\pdfoutput=1

\documentclass[conference,onecolumn,11pt]{IEEEtran}
\IEEEoverridecommandlockouts
\usepackage{amsmath,amssymb,amsfonts}
\usepackage{algorithmic}
\usepackage{graphicx}
\usepackage{textcomp}
\usepackage{xcolor}
\def\BibTeX{{\rm B\kern-.05em{\sc i\kern-.025em b}\kern-.08em
    T\kern-.1667em\lower.7ex\hbox{E}\kern-.125emX}}

\usepackage{xspace}
\usepackage[amsmath,thmmarks]{ntheorem}
\usepackage[hidelinks]{hyperref}
\usepackage[nameinlink,capitalize]{cleveref}

\usepackage{thmtools}
\usepackage{thm-restate}

\theoremstyle{plain}

\newtheorem{theorem}{Theorem}[section]
\newtheorem{lemma}[theorem]{Lemma}

\newtheorem{corollary}[theorem]{Corollary}
\newtheorem{claim}{Claim}
\newtheorem{example}[theorem]{Example}

\makeatletter
\@addtoreset{claim}{theorem}
\makeatother

\newenvironment{claimproof}{
  \begin{IEEEproof}}{\end{IEEEproof}}

\usepackage[numbers]{natbib}

\usepackage{mathtools}
\usepackage{caption}
\usepackage{subcaption}

\usepackage{tikz}
\usetikzlibrary{arrows,shapes, calc, fit, positioning,decorations.pathmorphing,decorations.pathreplacing}
\usepackage{pgfplots}

\usepackage[sc]{mathpazo}
\DeclarePairedDelimiter\floor{\lfloor}{\rfloor}
\DeclarePairedDelimiter\angles{\langle}{\rangle}

\newcommand{\ie}{i.\,e.}

\newcommand{\bigO}[1]{\ensuremath{\mathcal{O}(#1)}\xspace}
\newcommand{\N}{\ensuremath{\mathbb{N}}\xspace}
\newcommand{\Nplus}{\ensuremath{\mathbb{N}_{>0}}\xspace}
\newcommand{\Npos}{\Nplus}
\newcommand{\Z}{\ensuremath{\mathbb{Z}}\xspace}
\newcommand{\DTC}{\ensuremath{\textup{\textsf{DTC}}}\xspace}
\newcommand{\LRECeq}{\ensuremath{\textup{\textsf{LREC}}_{=}}\xspace}
\newcommand{\LREC}{\ensuremath{\textup{\textsf{LREC}}}\xspace}
\newcommand{\STC}{\ensuremath{\textup{\textsf{STC}}}\xspace}
\newcommand{\STCC}{\ensuremath{\textup{\textsf{STC+C}}}\xspace}
\newcommand{\FOC}{\ensuremath{\textup{\textsf{FO+C}}}\xspace}
\newcommand{\FO}{\ensuremath{\textup{\textsf{FO}}}\xspace}
\newcommand{\FP}{\ensuremath{\textup{\textsf{FP}}}\xspace}
\newcommand{\FPC}{\ensuremath{\textup{\textsf{FP+C}}}\xspace}
\newcommand{\C}[2]{\ensuremath{\textup{\textsf{C}}_{#1}^{#2}}\xspace} \newcommand{\WL}[2]{\ensuremath{\textup{\textsf{WL}}_{#1}^{#2}}\xspace} \newcommand{\LOGSPACE}{\ensuremath{\textup{\textsf{LOGSPACE}}}\xspace}
\newcommand{\PTIME}{\ensuremath{\textup{\textsf{PTIME}}}\xspace}
\newcommand{\NP}{\ensuremath{\textup{\textsf{NP}}}\xspace}
\newcommand{\Ck}[1]{\C{k}{#1}} \newcommand{\logCk}{\Ck{\bigO{\log n}}}
\newcommand{\Cplain}{\ensuremath{\textup{\textsf{C}}}\xspace}
\newcommand{\qd}{\ensuremath{\textup{\textsf{qd}}}}
\newcommand{\MC}{\ensuremath{\mathcal{M}}}

\newcommand{\stc}{\ensuremath{\textup{\textsf{{stc}}}}}
\newcommand{\lrec}{\ensuremath{\textup{\textsf{{lrec}}}}}
\newcommand{\lreceq}{\ensuremath{\textup{\textsf{{lrec}}}_=}}
\renewcommand{\phi}{\varphi}
\newcommand{\free}{\ensuremath{\textup{free}}}
\newcommand{\deff}{\coloneqq}

\newcommand{\Lor}{\bigvee}

\newcommand{\CP}{{\mathcal{P}}}
\newcommand{\CQ}{{\mathcal{Q}}}
\newcommand{\ceil}[1]{\left\lceil#1\right\rceil}

\newcommand{\set}[1]{\ensuremath{\left\{#1}\right\}\xspace}
\newcommand{\setc}[2]{\ensuremath{\left\{#1 \mid #2}\right\}\xspace}
\newcommand{\bigset}[1]{\ensuremath{\bigl\{#1}\bigr\}\xspace}
\newcommand{\bigmid}{\mathrel{\big|}}

\newcommand{\abs}[1]{\left\lvert#1\right\rvert}

\DeclareMathOperator{\wt}{wt}
\DeclareMathOperator{\mul}{mul}
\DeclareMathOperator{\awt}{awt}
\DeclareMathOperator{\amul}{amul}
\newcommand{\dagle}{\trianglelefteq}
\newcommand{\dagsle}{\triangleleft}

\definecolor{orange}{RGB}{230,159,0}             \definecolor{blue}{RGB}{86,180,233}           \definecolor{green}{RGB}{0,158,115}             \definecolor{marine}{RGB}{0,114,178}             \definecolor{red}{RGB}{213,94,0}              \definecolor{purple}{RGB}{204,121,167}          

\newcommand{\A}{\ensuremath{\mathcal{A}}\xspace}
\newcommand{\B}{\ensuremath{\mathcal{B}}\xspace}
 
\begin{document}
\title{Simulating Logspace-Recursion with Logarithmic Quantifier Depth}

\author{\IEEEauthorblockN{Steffen van Bergerem\IEEEauthorrefmark{1},
Martin Grohe\IEEEauthorrefmark{2},
Sandra Kiefer\IEEEauthorrefmark{3},
and Luca Oeljeklaus\IEEEauthorrefmark{2}\thanks{\IEEEauthorrefmark{1} This work was funded by the Deutsche Forschungsgemeinschaft (DFG, German Research Foundation) --- project number 431183758 (gefördert durch die Deutsche Forschungsgemeinschaft (DFG) --- Projektnummer 431183758).}
\thanks{\IEEEauthorrefmark{3} This research was supported by the Glasstone Benefaction, University of Oxford [Violette and Samuel Glasstone Research Fellowships in Science 2022].}
\thanks{\IEEEauthorrefmark{2} Funded by the European Union (ERC, SymSim,
101054974). Views and opinions expressed are however those of the
author(s) only and do not necessarily reflect those of the European
Union or the European Research Council. Neither the European Union
nor the granting authority can be held responsible for them.}}

\IEEEauthorblockA{\IEEEauthorrefmark{1}Humboldt-Universität zu Berlin, Berlin, Germany\\
Email: steffen.van.bergerem@informatik.hu-berlin.de}

\IEEEauthorblockA{\IEEEauthorrefmark{2}RWTH Aachen University, Aachen, Germany\\
Email: \{grohe,oeljeklaus\}@informatik.rwth-aachen.de}

\IEEEauthorblockA{\IEEEauthorrefmark{3}University of Oxford, Oxford, United Kingdom\\
Email: sandra.kiefer@cs.ox.ac.uk}
}

\IEEEoverridecommandlockouts
\IEEEpubid{\makebox[\columnwidth]{979-8-3503-3587-3/23/\$31.00~
\copyright2023 IEEE \hfill} \hspace{\columnsep}\makebox[\columnwidth]{ }}

\maketitle
\thispagestyle{plain}
\pagestyle{plain}

\begin{abstract}
The fixed-point logic \LRECeq was developed by Grohe et al.\ (CSL 2011) in the quest for a logic to capture all problems decidable in logarithmic space.
It extends \FOC, first-order logic with counting, by an operator that formalises a limited form of recursion.
We show that for every \LRECeq-definable property on relational structures,
there is a constant $k$ such that the \(k\)-variable fragment of first-order logic with counting quantifiers expresses the property via formulae of logarithmic quantifier depth.
This yields that any pair of graphs separable by the property can be distinguished with the $k$-dimensional Weisfeiler--Leman algorithm in a logarithmic number of iterations.
In particular, it implies that a constant dimension of the algorithm identifies every interval graph and every chordal claw-free graph in logarithmically many iterations,
since every such graph admits \LRECeq-definable canonisation.
\end{abstract}


\section{Introduction}
\label{sec:introduction}
By Fagin's celebrated theorem \cite{Fagin_1974}, over all finite structures, the complexity class \(\NP\) is precisely the class of all computational problems that can be expressed via formulae in existential second-order logic. This means that the problem to decide whether a structure has a certain property is in \(\NP\) if and only if there is a formula in existential second-order logic that defines the property. 
The theorem can be seen as the starting point of the field of descriptive complexity theory \cite{immerman1999descriptive,grohe2017descriptive}, which aims at describing or \emph{capturing} complexity classes via logics. Milestones include the results that, on ordered structures, fixed-point logic \(\FP\) captures \(\PTIME\) \cite{immerman1986immvardi,vardi1982immvardi} and deterministic transitive closure logic \(\DTC\) captures \(\LOGSPACE\) \cite{immerman1987dtc}. However, on general unordered structures, neither of these two results holds; for both complexity classes, the quest for capturing logics still continues and has possibly become the most important question in the field.

Concerning \(\LOGSPACE\), even adding counting operators to \(\DTC\) does not capture the class on trees yet \cite{etessami2000trees}. Towards tackling this, Grohe et al.\ designed in \cite{laubner2013recursion} the logic \(\LREC\), which captures \(\LOGSPACE\) on directed trees and is strictly contained in \(\FPC\), the extension of \(\FP\) by counting quantifiers. \(\LREC\) extends first-order logic with counting by an operator which enables a limited version of recursion. The idea behind it is that, as in fixed-point logics, some power of fixed-point operators should be allowed, but the amount of possible recursion shall not lead to the expressive power exceeding logarithmic-space computation.

By extending \(\LREC\) further to the logic \(\LRECeq\), a logic to capture \(\LOGSPACE\) on all undirected trees and on all interval graphs was found \cite{laubner2013recursion}. \(\LRECeq\) is strictly contained in \(\FPC\) \cite{dawar2022lrec} and in the logic Choiceless Logarithmic Space \cite{graedel2019choiceless}.

 More standard ``first-order'' logics such as transitive closure logic and its fragments quantify over vertices of the input graph. Addressing a single vertex in an $n$-element graph requires logarithmically many bits. Thus, to remain in \(\LOGSPACE\), such a logic can only store a bounded number of vertices at any time. This severely limits the expressiveness. 
The limited-recursion operator of \(\LREC\) as an explicit resource management allows it to use the logarithmic space in a more effective way and look at more vertices at the same time. An easy example illustrating how this might be possible is the following: suppose we have already stored a vertex $v$ of degree $d$. Then we only need $\log d$ bits to address any of its neighbours. As $d$ may be much smaller than $n$, this may allow us to save space. In this way, the logic \(\LREC\) is similar to choiceless polynomial time (\textsf{CPT}) \cite{BlassGS99,GradelG15}, 
which also has an explicit resource control, albeit to be able to more effectively exploit polynomial time. 
In the same way that \textsf{CPT} is strictly more powerful than \FPC and does not fit into the standard framework of finite-variable logics, \(\LREC\) is more powerful than deterministic and symmetric transitive closure logic with counting, and it is not even contained in full transitive closure logic with counting, a logic capturing nondeterministic logarithmic space on ordered structures. Larger complexity classes such as $\textsf{AC}^1$, which describes parallel logarithmic-time complexity, are naturally described in terms of fixed-point logic with logarithmically many iterations. While it is known that \(\LREC\) is contained in \FPC, it is not obvious that it can be simulated by logarithmically many fixed-point iterations, as the limited recursion of \(\LREC\) may be polynomially deep. This is the question we address in this paper.

To understand the expressiveness of fixed-point logic with counting, it has turned out to be very fruitful to embed it into the finite-variable logics of first-order logic with counting \cite{cai1992optimal,Otto97}. Then the number of fixed-point iterations corresponds naturally to the quantifier depth.

Over the decades of research, these counting-logic fragments have exhibited links to many other areas from practical and theoretical computer science \cite{atserias2013sherali,atserias2018proofs,cai1992optimal,dell2018lovasz,grohe2015pebbles,grohe2021logicgnn,morris2019neural}. 
A striking connection exists to the Weisfeiler--Leman algorithm, which is a procedure that computes and refines in an iterative and isomorphism-invariant way colours in the input graph. For every $k \in \Npos$, its $k$-dimensional version runs in polynomial time and the computed colourings can often be used to detect non-isomorphism of graphs. As it turns out, the $k$-dimensional Weisfeiler--Leman algorithm ($k$-$\WL{}{}$) is just as expressive as the logic $\C{k+1}{}$, the $(k+1)$-variable fragment of first-order logic enriched with counting quantifiers: it computes distinct colourings on two input graphs if and only if there is a distinguishing formula in $\C{k+1}{}$ for them. Moreover, the number of iterations of the algorithm needed to obtain distinct colourings corresponds to the quantifier depth of a distinguishing formula. By this correspondence, from a perspective of descriptive complexity theory, both the dimension of the algorithm and the number of iterations that it needs to produce an output are parameters worth being studied.

Concerning the dimension of the algorithm that is needed to distinguish two graphs, over the past years, many new insights have been obtained, also exploiting the link to counting logics. For example, forests can be identified with $1$-$\WL{}{}$, interval graphs with $2$-$\WL{}{}$ \cite{evdokimov2000interval}, and planar graphs with $3$-$\WL{}{}$ \cite{kiefer19planar}. Also for many other natural graph classes, bounds on the necessary dimension to tackle the isomorphism problem in the class are known \cite{grohe2019genus,grohe2019rank,grohe2000embeddable,kiefer2022decompositions}.

Concerning the number of iterations, much less is known. Fürer proved a linear lower bound on the number of iterations of $k$-$\WL{}{}$ \cite{fuerer2001lower}, which was improved to $n^{\Omega(k/log k)}$ \cite{berkholz2016near} and recently to $n^{\Omega(k)}$ \cite{GroheLN23} on $k$-ary relational structures. As to upper bounds, for $k=2$, first progress over the trivial upper bound of $\Theta(n^k)$ has been made in \cite{kiefer2019iteration}, and the best known upper bound is $O(n\log n)$ on graphs of order $n$ \cite{lichter2019iteration}. This has been generalised to a bound of $O(n^{k-1}\log n)$ for all $k\ge 2$ in \cite{GroheLN23}. The number of iterations is crucial for the parallelisability of the algorithm. For $\ell \geq \log n$, it holds that $\ell$ iterations of $k$-$\WL{}{}$ can be simulated in $\bigO{\ell}$ steps on a PRAM with $O(n^k)$ processors. This implies that if  $k$-$\WL{}{}$ distinguishes all pairs of graphs of order $n$ in a class $\mathcal{C}$ in $\bigO{\log n}$ iterations, then deciding isomorphism for $\mathcal{C}$ is in the complexity class \textsf{TC}$^1$. This is the case for all graph classes of bounded treewidth and all maps \cite{grohe2006pebbles} as well as all planar graphs \cite{grohe2021logarithmic}. 

In this paper, we extend the result to all classes of interval graphs and, as a by-product, we obtain the same for chordal claw-free graphs.

\paragraph*{Our results}

We study the expressive power of the Weisfeiler--Leman algorithm when restricted to a logarithmic number of iterations. This restriction was first introduced as a means of showing that the graph isomorphism problem for graphs of bounded treewidth is in \textsf{TC}$^1$ \cite{grohe2006pebbles}. In fact, we transcend to the logical perspective on the algorithm and prove that for every property on relational structures that is definable in the logic \(\LRECeq\), there is a number $k \in \mathbb{N}$ such that the logic \(\Ck{}\) expresses the property via a family of formulae of logarithmic quantifier depth, which is equivalent to $(k-1)$-$\WL{}{}$ detecting the property in a logarithmic number of iterations. That is, intuitively speaking, we can simulate logspace recursion with a logarithmic number of iterations of a suitable dimension of the Weisfeiler--Leman algorithm or, equivalently, with a logarithmic quantifier depth in \(\Ck{}\).

The formal statement of the result is as follows.

\begin{restatable}{theorem}{lreqeqtock} \label{theo:lreceq_to_c_k}
  For every vocabulary \(\tau\)
  and every \(\LRECeq[\tau]\)-formula \(\phi(\bar{x}, \bar{\kappa})\),
  there is a constant \(k \in \N\) such that for every \(n \in \N\),
  there is a family of \(\logCk\)-formulae
  \(\bigl(\psi_{\bar{m}}(\bar{x})\bigr)_{\bar{m} \in [n]^{\abs{\bar{\kappa}}}}\)
  such that for all \(\tau\)-structures \(\A\) of size \(\abs{\A} \leq n\),
  all \(\bar{v} \in \bigl(V(\A)\bigr)^{\abs{\bar{x}}}\),
  and all \(\bar{m} \in [\abs{\A}]^{\abs{\bar{\kappa}}}\), it holds that
  \[\A \models \phi(\bar{v}, \bar{m}) \iff \A \models \psi_{\bar{m}}(\bar{v}).\]
\end{restatable}

In the proof, we restructure the recursive computation of the \(\LRECeq\)-operator to obtain a computation tree of logarithmic height and small bag overlap. The construction of the desired formulae is then similar to the approach presented in \cite{grohe2021logarithmic}: we build the formula from bottom to top along the tree decomposition, resulting in logarithmic quantifier depth. Here, we need to take care that the number of variables really stays constant.

As an example for the usefulness of the result, we then apply the result to the class of all interval graphs, which is the class of graphs which initially motivated us to study the power of the logarithmic Weisfeiler--Leman algorithm. The class of interval graphs is relevant in many application areas, for example in biology \cite{zhang1994algorithm} and in operations research \cite{darmann2010resource}, and many usually computationally hard problems are known to be tractable on them \cite{gupta1982efficient, keil1985finding}.
Köbler et al.\ \cite{koebler2011interval} gave a \(\LOGSPACE\)-algorithm
for the isomorphism problem on interval graphs.
Their result, however, is purely algorithmic and does not translate to results in terms of logics, making it incomparable to our theorem. Since by \cite{laubner2013recursion}, for every interval graph, there is an \(\LRECeq\)-formula that identifies it, we can deduce that a constant number of variables suffices to identify every interval graph with a \(\Cplain\)-sentence of logarithmic quantifier depth. 

\begin{restatable}{theorem}{interval}\label{theo:interval}
  There is a $k \in \N$ such that for every $n \in \N$ and interval graph $G$ of order $n$,
  there is a formula $\Phi^G \in \logCk$ that describes $G$ up to isomorphism.
\end{restatable}

As a by-product, using \cite{grussien2019capturing}, we obtain an analogous statement for the class of chordal claw-free graphs.

\begin{restatable}{theorem}{clawfreechordal} \label{theo:claw_free_chordal}
  There is a $k \in \N$ such that for every chordal claw-free graph $G$ of order $n$,
  there is a formula $\Phi^G \in \logCk$ that describes $G$ up to isomorphism.
\end{restatable}

Afterwards, we analyse interval graphs in more detail and sketch a second, direct proof to show that isomorphism types of those graphs are definable in $\Cplain$ with a constant number of variables and logarithmic quantifier depth. 
The proof avoids translating \(\LRECeq\)-formulae and proceeds straight via a decomposition of the graph.
 
\section{Preliminaries}
\label{sec:preliminaries}
We denote a tuple of elements $(x_1, \dots, x_k)$ as $\bar{x}$. Two tuples $\bar{x} = (x_1,\dots, x_k)$, $\bar{y} = (y_1, \dots, y_\ell)$ are said to be \emph{compatible} if $k = \ell$ and hence $|\bar{x}| = |\bar{y}|$. We refer to the $i$th position of a tuple $\bar{x}$ as $\bar{x_i}$.
For all $k,\ell \in \N$, we define $[k,\ell] \deff \setc{i \in \N}{k \leq i \leq \ell}$ and $[k] \coloneqq [1,k]$.

\subsection{Structures}

A \emph{vocabulary} is a non-empty finite set of relation symbols.
Each symbol \(R \in \tau\) has a fixed \emph{arity} \(a_R \in \N\).
A \(\tau\)\emph{-structure} \A consists of a \emph{domain} --- a non-empty, finite set \(V(A)\) --- and,
for each \(R \in \tau\), a relation \(R(\A) \subseteq V(\A)^{a_R}\).
By the \emph{order} of a structure, we refer to its cardinality $|\A|$.
We may write $a_1\dots a_k \in R(\A)$ instead of $(a_1,\dots,a_k) \in R(\A)$.
An \emph{isomorphism} between $\tau$-structures $\A$ and $\B$ is a relation-preserving bijection $\mu \colon V(\A) \to V(\B)$,
\ie, for all $k$-ary $R \in \tau$, and all $a_1,\dots,a_k \in V(\A)$,
it must hold that $(a_1,\dots,a_k) \in R(\A) \iff (\mu(a_1),\dots,\mu(a_k)) \in R(\B)$.
We then call $\A$ and $\B$ \emph{isomorphic}, denoted as $\mu \colon \A \cong \B$.
We may omit $\mu$ if the particular mapping does not interest us.

For a $\tau$-structure $\A$, we define the \emph{two-sorted structure} $\A^+$ extending $\A$ as
\[\A^+ \coloneqq \left(V(\A),\set{R(\A)}_{R\in\tau},N(\A),\le,S,\min,\max\right),\]
where $N(\A) \coloneqq \set{0,\ldots,|\A|}$, $\le$ is the corresponding linear order, $\min$, $\max$ are unary singleton relations defining the minimum and maximum of $\le$, and $S$ is the binary successor relation.
Every \emph{domain variable} then ranges over the universe $V(\A)$ and is from the set $x_1,x_2,\dots$, whereas every \emph{number variable} ranges on $N(\A)$ and is from the set $\iota_1, \iota_2 \dots$. We may deviate from this convention and use the symbols $x,y,z \dots$ for domain, resp.\ $\iota,\kappa,\lambda \dots$ for number variables. However, these should be understood to be placeholders for values from $x_1,\dots$ resp.\ $\iota_1,\dots$ used in an effort towards enhanced readability.
To represent elements of $V(\A)$ resp.\ $N(\A)$, we employ symbols from $u,v,w,\dots$ resp.\ $i,j,p,q,\dots$.

Two two-sorted structures are isomorphic if their underlying $\tau$-structures are isomorphic.

Although being limited to an initial segment of the natural numbers, we can represent larger numbers through tuples $\bar{i} \in N(\A)^k$, which are then interpreted as base-$(|\A|+1)$ numbers as
\[\langle\bar{i}\rangle_\A \coloneqq \sum_{j=1}^k \bar{i}_j \cdot (|\A|+1)^{j-1}.\]

We define an \emph{interpretation} (or \emph{assignment}) to be a mapping $\alpha$ assigning, to each domain variable $x_i$ a value in $V(\A)$ and to each number variable $\iota_i$ a value in $N(\A)$. 
Since we consider interpretations only together with concrete formulae, it is sufficient if every variable occurring in the formula is assigned a value.
Given domain resp.\ number variable tuples $\bar{x}$ and $\bar{\iota}$ as well as compatible domain resp.\ number tuples $\bar{v} \in V(\A)^k$ and $\bar{p} \in N(\A)^\ell$, we write $\alpha[\bar{v}/\bar{x},\bar{p}/\bar{\iota}]$ to mean the assignment $\alpha$ modified to the effect that, for all $i \in [k]$ and $j \in [\ell]$, $\bar{x_i}$ is assigned the value $\bar{v_i}$ and $\bar{\iota_j}$ is assigned the value $\bar{p_j}$.

We call a variable \emph{bound} if it occurs within the scope of a corresponding quantifier. Otherwise, we call it \emph{free}. In particular, given a formula $\phi$, we write $\phi(x_1,\dots,x_k)$ to express that $\free(\phi) \subseteq \set{x_1,\dots,x_k}$ are distinct and that they are those variables that may occur free within $\phi$. Those formulae in which all variables occur bound are \emph{sentences}.

\subsection{Graphs}

A \emph{(directed) graph} is an $\set{E}$-structure $G \coloneqq (V(G),E(G))$ over a domain of \emph{vertices} $V(G)$ and a binary \emph{edge} relation $E(G)$. The \emph{order} of $G$ is $\abs{G} \coloneqq |V(G)|$.

A graph is \emph{undirected} if $E(G)$ is symmetric and irreflexive, in which case we consider the elements $vw \in E(G)$ to be unordered.
For such $G$ and $v \in V(G)$, we denote by $N_G(v) \coloneqq \set{w \in V(G) \mid vw \in E(G)}$ resp.\ $N_G[v] \coloneqq N_G(v) \cup \set{v}$ the \emph{open} resp.\ \emph{closed} neighbourhood of $v$ in $G$.
Letting $W \subseteq V(G)$, we define the \emph{subgraph induced by} $W$ \emph{in} $G$ as $G[W] \coloneqq (W,\set{vw\in E(G) \mid v,w \in W})$.

A directed graph is \emph{acyclic} if there is no $k \in \N$ for which there exists a sequence of distinct edges $(v_1,v_2),(v_2,v_3)\dots,(v_{k-1},v_k)$ with $v_i \neq v_j$ for distinct $i,j \in [k-1]$ such that $v_1 = v_k$.
We refer to \emph{directed acyclic graphs} as \emph{DAG}s.
The \emph{height} of a DAG is the length of a longest path in it. 

Now, let \(G\) be a directed graph of order \(n \deff \abs{G}\).
For every \(v \in V(G)\),
we let \(N_G^+(v) \deff \setc{w \in V(G)}{vw \in E(G)}\)
and \(N_G^-(v) \deff \setc{u \in V(G)}{uv \in E(G)}\)
be the sets of \emph{out-neigh\-bours} and \emph{in-neighbours} of \(v\),
and \(\deg_G^+(v) \deff \abs{N_G^+(v)}\)
and \(\deg_G^-(v) \deff \abs{N_G^-(v)}\)
be the \emph{out-degree} and the \emph{in-degree} of \(v\).

Let \(\dagle_G\) denote the reflexive transitive closure of \(E(G)\).
In these and similar notations,
we omit the subscript \({}_G\) if the graph \(G\) is clear from the context.
We call a node of out-degree \(0\) a \emph{leaf} of \(G\).
We call \(G\) \emph{rooted} if \(\dagle\) has a unique minimal element \(r\)
that we call the \emph{root} of \(G\).
Note that every \(v \in V(G)\) is reachable from \(r\).

\subsection{Logics}

This section presumes familiarity with first-order logic (\FO) and standard model-theoretic notation,
suitable overviews for which can be found in \cite{ebbinghaus2005finite,libkin2004elements}.
Given a formula $\phi$, a structure $\A$ and an assignment $\alpha$, we write $(\A,\alpha) \models \phi$ to express that the formula $\phi$ is \emph{true} given $\A$ and having assigned the variables as in $\alpha$.
In particular, we use the shorthands $\top \coloneqq \forall x \, (x = x)$ and $\bot \coloneqq \neg \top$.

Let us introduce an extension of \FO,
\emph{first-order logic with counting} (\FOC),
which operates on the two-sorted structures described above (see also \cite{grohe2017descriptive}).
It extends \FO over a second domain and adds two quantifiers, which we now define recursively.
In the remainder of the section, let $\tau$ be an arbitrary vocabulary.
Further, let $\A^+$ be a two-sorted $\tau$-structure,
let $\phi$ be an $\FOC[\tau]$-formula,
$\alpha$ an interpretation and $\iota$ be a number variable.
Then $\exists \iota \, \varphi$ is an $\FOC[\tau]$-formula, and it is satisfied by $(\A^+,\alpha)$ iff
\[
\set{i \in N(\A)\mid (\A^+,\alpha[i/\iota]) \models \phi} \neq \varnothing
\]
holds.

\FOC also adds the $\#$-quantifier, evaluating to some natural number.
Let $x$ be a domain variable and $\kappa$ an additional number variable.
Then $\# x \, \phi = \kappa$ and $\# \iota \, \phi = \kappa$ are $\FOC[\tau]$-formulae and
\begin{align*}
                    & \A^+ \models \# x \, \phi = \kappa \\
    \iff & \left|\set{v \in V(\A) \mid (\A^+,\alpha[v/x]) \models \phi}\right| = \alpha(\kappa)
\end{align*}
resp.
\begin{align*}
                    & \A^+ \models \# \iota \, \phi = \kappa \\
    \iff & \left|\set{i \in N(\A) \mid (\A^+,\alpha[i/\iota]) \models \phi}\right| = \alpha(\kappa).
\end{align*}

Given a formula $\phi(\bar{x},\bar{\iota})$, we define the set of domain/number tuples satisfying $\phi$ as 
\begin{align*}
    & \phi[\A^+,\alpha;\bar{x},\bar{\iota}] \coloneqq \\ 
    & \set{\bar{v}\bar{i} \in V(\A)^{|\bar{x}|} \times N(\A)^{|\bar{\iota}|}\mid (\A^+,\alpha[\bar{v}/\bar{x},\bar{i}/\bar{\iota}]) \models  \phi(\bar{x},\bar{\iota})}.
\end{align*}

We now turn towards describing \LREC and \LRECeq,
two extensions of \FOC which were first introduced in \cite{laubner2013recursion} in the quest for a logic capturing \textsf{LOGSPACE}.
The set of all $\LREC[\tau]$-formulae is obtained by extending the syntax of $\FOC[\tau]$ by the following rule.
Let $\bar{x},\bar{y}_1,\bar{y}_2$ be compatible domain variable $k$-tuples,
and $\bar{\iota},\bar{\kappa}$ be non-empty number variable tuples.
Then, if $\phi_\texttt{E}, \phi_\texttt{C}$ are $\LREC[\tau]$-formulae,
\[\phi \coloneqq [\lrec_{\bar{y}_1,\bar{y}_2,\bar{\iota}} \phi_\texttt{E}, \phi_\texttt{C}](\bar{x},\bar{\kappa})\]
is an $\LREC[\tau]$-formula with
$\free(\phi) \coloneqq (\free(\phi_\texttt{E}) \setminus (\bar{y}_1 \cup \bar{y}_2)) \cup (\free(\phi_\texttt{C}) \setminus (\bar{y}_1 \cup \bar{\iota})) \cup \bar{x} \cup \bar{\kappa}$.
Given a two-sorted $\tau$-structure $\A^+$ and an assignment $\alpha$,
the formula $\phi$ recursively defines a relation $X \subseteq V(\A)^{k} \times \N$ such that
\[(\A^+,\alpha) \models \phi \iff (\alpha(\bar{x}),\langle\alpha(\bar{\kappa})\rangle_\A) \in X.\]
We now describe how $X$ is obtained. 
Initially, define a graph $\texttt{G} \coloneqq (\texttt{V}, \texttt{E})$
with \(\texttt{V} \coloneqq V(\A)^k\)
and \(\texttt{E} \coloneqq \phi_{\texttt{E}}[\A,\alpha; \bar{y}_1,\bar{y}_2])\).
That is, $\texttt{G}$ is a directed graph on the $k$-tuples of $V(\A)$,
and the edges are precisely those pairs \((\bar{v}_1, \bar{v}_2)\)
with $\bar{v}_1,\bar{v}_2 \in V(\A)^k$
such that $(\A,\alpha[\bar{v}_1/\bar{y}_1,\bar{v}_2/\bar{y}_2]) \models \phi_{\texttt{E}}(\bar{y}_1,\bar{y}_2)$.
Over $\texttt{G}$, the formula $\phi_{\texttt{C}}$ then defines a vertex labelling
\[\texttt{C}(\bar{v}_1) \coloneqq \set{\langle \bar{i} \rangle_\A \mid \bar{i} \in \phi_{\texttt{C}}[\A^+,\alpha[\bar{v}_1/\bar{y}_1];\bar{\iota}]}.\]
A tuple $(\bar{v},\bar{i})$ with $\bar{v} \in V(\A)^k$ and a ``resource term'' $\bar{i} \in N(\A)^{|\iota|}$ is contained in $X$ if $\angles{i}_\A > 0$ (that is, there are still resources left) and
\[
\left|\set{\bar{w} \in N^+_{\texttt{G}}(\bar{v}) \Bigg| \left(\bar{w},\floor*{\frac{\angles{i}_\A-1}{\deg^-_{\texttt{G}}(\bar{w})}}\right) \in X}
\right| \in \texttt{C}(\bar{v}).
\]

The logic \LRECeq replaces the $\lrec$-operator by the $\lreceq$-operator, which allows the definition of an equivalence relation on the constructed graph.
We recall the above defined structure, assignment, and variable tuples. Then, letting  $\phi_{\texttt{=}},\phi_{\texttt{E}},\phi_{\texttt{C}}$, be $\LRECeq[\tau]$-formulae, we obtain a new $\LRECeq[\tau]$-formula

\[\phi \coloneqq [\lrec_{\bar{y}_1,\bar{y}_2,\bar{\iota}} \phi_{\texttt{=}},\phi_{\texttt{E}},\phi_{\texttt{C}}](\bar{x},\bar{\kappa}).\]

The semantics are a bit more complex. First, we construct a graph $\texttt{G'}$ as before.
Then, letting $\sim$ be the equivalence relation defined by $\phi_{\texttt{=}}[\A^+,\alpha;\bar{y}_1,\bar{y}_2]$ on $\texttt{V'}$, we define a new graph
\begin{align*}
    \texttt{G} \coloneqq \big(&\texttt{V} \coloneqq \texttt{V'}/_\sim,\\ &\texttt{E} \coloneqq \set{(\bar{v}_{1/\sim},\bar{v}_{2/\sim})\in \texttt{V}^2\mid (\bar{v}_1,\bar{v}_2) \in \texttt{E'}}\big)
\end{align*}
contracting the vertices from $\texttt{V'}$ into their equivalence classes while maintaining the edges.
For all $\bar{v}_{1/\sim} \in \texttt{V}$, we define
\[\texttt{C}(\bar{v}_{1/\sim}) \coloneqq \set{\langle \bar{i} \rangle_\A \mid \exists \bar{v}' \in \bar{v}_{1/\sim}: \bar{i} \in \phi_{\texttt{C}}[\A^+,\alpha[\bar{v}'/\bar{y}_1];\bar{\iota}]}.\]
The relation $X$ is then defined as previously. 

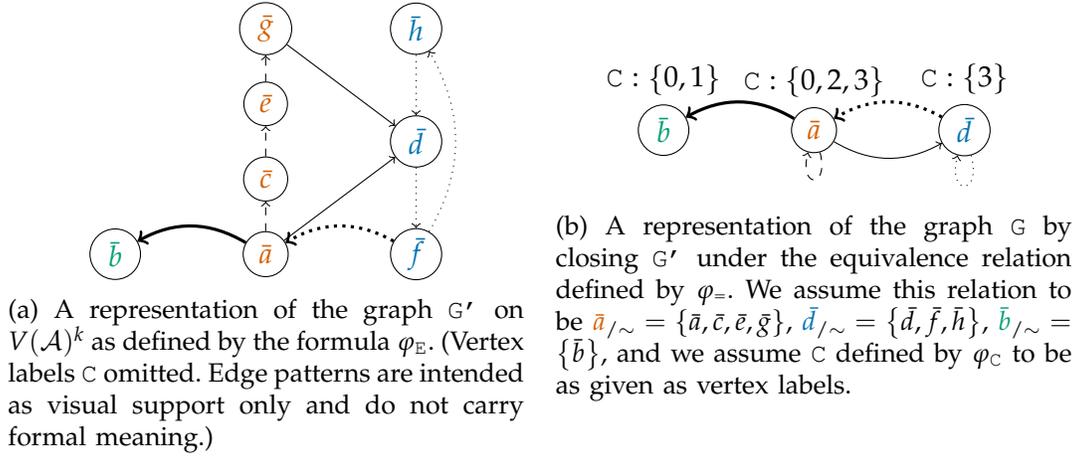
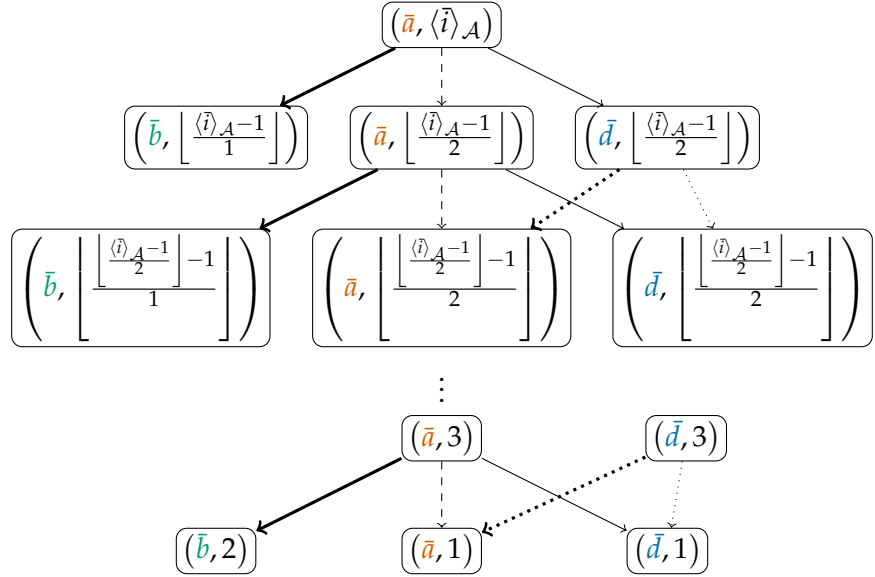
\begin{figure*}
    \centering
    \begin{tabular}[b]{c}
    \begin{tabular}[b]{cc}
      \begin{subfigure}{0.44\textwidth}
      \centering
        \begin{tikzpicture}
            \node[circle,draw=black,inner sep=3pt] (a) at (0,0) {$\textcolor{red}{\bar{a}}$};
            \node[circle,draw=black,inner sep=3pt] (b) at (-2,0) {$\textcolor{green}{\bar{b}}$};
            \node[circle,draw=black,inner sep=3pt] (c) at (0,1) {$\textcolor{red}{\bar{c}}$};
            \node[circle,draw=black,inner sep=3pt] (d) at (2,1.5) {$\textcolor{marine}{\bar{d}}$};
            \node[circle,draw=black,inner sep=3pt] (e) at (0,2) {$\textcolor{red}{\bar{e}}$};
            \node[circle,draw=black,inner sep=2pt] (f) at (2,0) {$\textcolor{marine}{\bar{f}}$};
            \node[circle,draw=black,inner sep=3pt] (g) at (0,3) {$\textcolor{red}{\bar{g}}$};
            \node[circle,draw=black,inner sep=3pt] (h) at (2,3) {$\textcolor{marine}{\bar{h}}$};
            
            \draw[->,very thick] (a) to[bend right] (b);
            \draw[->,dashed] (a) -- (c);
            \draw[->] (a) -- (d);
            \draw[->,dashed] (c) -- (e);
            \draw[->,dotted] (d) -- (f);
            \draw[->,dashed] (e) -- (g);
            \draw[->] (g) -- (d);
            \draw[->,very thick, dotted] (f) to[bend right] (a);
            \draw[->,dotted] (f) to[bend right] (h);
            \draw[->,dotted] (h) -- (d);

            \draw[white] (b) to[loop left] (b);
            \draw[white] (d) to[loop right] (d);
        \end{tikzpicture}
        \caption{A representation of the graph $\texttt{G'}$ on $V(\A)^k$ as defined by the formula $\phi_{\texttt{E}}$. (Vertex labels $\texttt{C}$ omitted. Edge patterns are intended as visual support only and do not carry formal meaning.)}
        \label{fig:lrec_ex_1}
    \end{subfigure} &
      \begin{subfigure}{0.44\textwidth}
            \centering
        \begin{tikzpicture}
                    \node[circle,draw=black,inner sep=3pt,label={above: $\texttt{C}: \set{0,2,3}$}]
                    (a) at (0,0) {$\textcolor{red}{\bar{a}}$};
                    \node[circle,draw=black,inner sep=3pt,label={above: $\texttt{C}: \set{0,1}$}]
                    (b) at (-2,0) {$\textcolor{green}{\bar{b}}$};
                    \node[circle,draw=black,inner sep=3pt,label={above: $\texttt{C}: \set{3}$}]
                    (d) at (2,0) {$\textcolor{marine}{\bar{d}}$};

                    \draw[->,very thick] (a) to[bend right] (b);
                    \draw[white] (b) to[loop left] (b);
                    \draw[->,dashed] (a) to[loop below] (a);
                    \draw[->,dotted] (d) to[loop below] (d);
                    \draw[white] (d) to[loop right] (d);
                    \draw[->] (a) to[bend right] (d);
                    \draw[->,very thick, dotted] (d) to[bend right] (a);
            \end{tikzpicture}
        \caption{A representation of the graph $\texttt{G}$ by closing $\texttt{G'}$ under the equivalence relation defined by $\phi_{\texttt{=}}$.
        We assume this relation to be $\textcolor{red}{\bar{a}}_{/\sim} = \set{\bar{a},\bar{c},\bar{e},\bar{g}}$, $\textcolor{marine}{\bar{d}}_{/\sim} = \set{\bar{d},\bar{f},\bar{h}}$, $\textcolor{green}{\bar{b}}_{/\sim} = \set{\bar{b}}$,
        and we assume $\texttt{C}$ defined by $\phi_\texttt{C}$ to be as given as vertex labels.}
        \label{fig:lrec_ex_2}
    \end{subfigure}
    \end{tabular}
    \\
    \begin{subfigure}{0.9\textwidth}
          \centering
        \begin{tikzpicture}
                    \node[] (x) at (0,10.5) {};
                    \node[draw=black,inner sep=2pt,rounded corners]
                    (1) at (0,10) {$\left(\textcolor{red}{\bar{a}}\phantom{\mathclap{\bar{b}}},\angles{\bar{i}}_{\A}\right)$};

                    \node[draw=black,inner sep=2pt,rounded corners]
                    (11) at (0,8.5) {$\left(\textcolor{red}{\bar{a}}\phantom{\mathclap{\bar{b}}},\floor*{\frac{\angles{\bar{i}}_\A-1}{2}}\right)$};
                    \node[draw=black,inner sep=2pt,rounded corners]
                    (12) at (-3,8.5) {$\left(\textcolor{green}{\bar{b}},\floor*{\frac{\angles{\bar{i}}_{\A}-1}{1}}\right)$};
                    \node[draw=black,inner sep=2pt,rounded corners]
                    (13) at (3,8.5) {$\left(\textcolor{marine}{\bar{d}},\floor*{\frac{\angles{\bar{i}}_\A-1}{2}}\right)$};

                    \draw[->,dashed] (1) to (11);
                    \draw[->,very thick] (1) to (12);
                    \draw[->] (1) to (13);

                    \node[draw=black,inner sep=2pt,rounded corners]
                    (21) at (0,6.5) {$\left(\textcolor{red}{\bar{a}}\phantom{\mathclap{\bar{b}}},\floor*{\frac{\floor*{\frac{\angles{\bar{i}}_\A-1}{2}}-1}{2}}\right)$};
                    \node[draw=black,inner sep=2pt,rounded corners]
                    (22) at (-4,6.5) {$\left(\textcolor{green}{\bar{b}}\phantom{\mathclap{\bar{b}}},\floor*{\frac{\floor*{\frac{\angles{\bar{i}}_\A-1}{2}}-1}{1}}\right)$};
                    \node[draw=black,inner sep=2pt,rounded corners]
                    (23) at (4,6.5) {$\left(\textcolor{marine}{\bar{d}}\phantom{\mathclap{\bar{b}}},\floor*{\frac{\floor*{\frac{\angles{\bar{i}}_\A-1}{2}}-1}{2}}\right)$};

                    \draw[->,dashed] (11) to (21);
                    \draw[->,very thick, dotted] (13) to (21);
                    \draw[->,very thick] (11) to (22);
                    \draw[->] (11) to (23);
                    \draw[->,dotted] (13) to (23);

                    \node[draw=black,inner sep=2pt,rounded corners]
                    (31) at (0,4.5) {$\left(\textcolor{red}{\bar{a}}\phantom{\mathclap{\bar{b}}},3\right)$};
                    \node[]
                    (31x) at (0,5.25) {$\vdots$};
                    \node[draw=black,inner sep=2pt,rounded corners]
                    (33) at (3.25,4.5) {$\left(\textcolor{marine}{\bar{d}},3\right)$};

                    \node[draw=black,inner sep=2pt,rounded corners]
                    (41) at (0,3) {$\left(\textcolor{red}{\bar{a}}\phantom{\mathclap{\bar{b}}},1\right)$};
                    \node[draw=black,inner sep=2pt,rounded corners]
                    (42) at (-3,3) {$\left(\textcolor{green}{\bar{b}}\phantom{\mathclap{\bar{b}}},2\right)$};
                    \node[draw=black,inner sep=2pt,rounded corners]
                    (43) at (3,3) {$\left(\textcolor{marine}{\bar{d}},1\right)$};

                    \draw[->,dashed] (31) -- (41);
                    \draw[->,very thick, dotted] (33) to (41);
                    \draw[->,very thick] (31) to (42);
                    \draw[->] (31) to (43);
                    \draw[->,dotted] (33) to (43);
            \end{tikzpicture}
            \caption{A visualisation of the DAG resulting from the recursive unfolding of the graph $\texttt{G}$ with respect to the parameters $(\bar{a},\bar{i})$.
            For every recursive step, the resource term
            (\ie, $\angles{\bar{i}}_\A$, $\floor*{\frac{\angles{\bar{i}}_\A-1}{2}}$, etc.)
            can be understood to be ``split'' equitably among the in-neighbours of the vertex in \(\texttt{G}\).
            Together with the requirement that the resource term must be positive, this ensures a logarithmic space bound.}
        \label{fig:lrec_ex_3}
    \end{subfigure}
  \end{tabular}
    \caption{A visualisation of the computation of the relation \(X\)
    conducted when evaluating an $\LRECeq[\tau]$-formula
    \(\phi \coloneqq [\lrec_{\bar{y}_1,\bar{y}_2,\bar{\iota}}
      \phi_{\texttt{\textup{=}}},
      \phi_{\texttt{\textup{E}}},
      \phi_{\texttt{\textup{C}}}](\bar{a},\bar{i})\).
    Let \(\A^+\) be a two-sorted \(\tau\)-structure.
    We assume the graph \(\texttt{G'}\), defined via the formula \(\phi_{\texttt{\textup{E}}}\),
    to be as shown in \cref{fig:lrec_ex_1}.
    We then assume the formula \(\phi_{\texttt{\textup{=}}}\)
    to yield the graph \(\texttt{G}\)
    and the formula \(\phi_{\texttt{\textup{C}}}\)
    to yield the vertex labelling \(\texttt{C}\)
    as shown in \cref{fig:lrec_ex_2}.
    Using \cref{fig:lrec_ex_3}, we now describe how the DAG is evaluated
    and thereby how the relation $X$ is computed, in a bottom-to-top fashion.
    A leaf, say $(\textcolor{green}{\bar{b}},2)$, is part of the relation $X$
    if $0 \in \texttt{\textup{C}}(\textcolor{green}{\bar{b}})$,
    since the criterion is whether the number of $(\textcolor{green}{\bar{b}},2)$'s children which are in $X$ (that is, $0$)
    occurs in $\texttt{\textup{C}}(\textcolor{green}{\bar{b}})$.
    Therefore, here it holds that $(\textcolor{green}{\bar{b}},2),(\textcolor{red}{\bar{a}},1) \in X$,
    but $(\textcolor{marine}{\bar{d}},1) \not \in X$.
    Then their predecessor $(\textcolor{red}{\bar{a}},3)$ is in $X$ since 
    $\abs{\set{(\bar{b}, 2), (\bar{a}, 1)}} = 2 \in \texttt{\textup{C}}(\textcolor{red}{\bar{a}})$,
    whereas $(\textcolor{marine}{\bar{d}},3) \not\in X$
    because $\abs{\set{(\bar{a}, 1)}} = 1 \not\in \texttt{\textup{C}}(\textcolor{marine}{\bar{d}})$.
    This then continues up to the root $(\textcolor{red}{\bar{a}},\angles{\bar{i}}_\A)$,
    which is contained in $X$ if the number $q$ of its children in $X$
    occurs in $\texttt{\textup{C}}(\textcolor{red}{\bar{a}})$.
    Finally, $\phi$ is satisfied iff the root is in $X$.}
    \label{fig:lrec_ex}
\end{figure*}
 See \cref{fig:lrec_ex} for an example of how the relation \(X\) is computed.
Readers wishing to develop a more in-depth understanding of \LREC and \LRECeq may want to look into \cite{laubner2013recursion}, which also includes concrete examples of properties that can be expressed in these logics.

We now move towards defining the logic $\Cplain$, \emph{first-order logic with counting quantifiers}, as the syntactical extension of $\FO$ by $\Cplain$\emph{-quantifiers} of the form
$\exists^{\ge n} x\, \phi(x)$ (\emph{there are at least $n$ elements $x$ satisfying $\phi(x)$}) for all $n\in \N$, which immediately yields the related quantifiers $\exists^{\le n} x\, \phi(x) \equiv \exists x\, \phi(x) \land \neg \exists^{\ge n+1} x\, \phi(x)$ and $\exists^{= n} x\, \phi(x) \equiv \exists^{\le n} x\, \phi(x) \land \exists^{\ge n} x\, \phi(x)$.

Before continuing, it is worth noting that \Cplain and \FOC are two distinct, separate logics, which happen to be similarly named. We attempt to shortly clarify their differences to preempt any confusion. The logic \Cplain is only a syntactical extension of \FO on relational structures, whereas \FOC is defined on two-sorted structures and has some access to quantification over the natural numbers.
For example, whether a graph is regular or not can be expressed by the \FOC-formula
\[\phi^{\FOC}_\text{regular} \coloneqq \exists \iota \, \big[\forall x \# y \, E(x,y) = \iota\big],\]
which can be understood as ``there exists a number $\iota$ such that every vertex $x$ has exactly $\iota$ neighbours''. On the other hand, as a consequence of the hardcoded aspect of numbers in \Cplain-quantifiers, any \Cplain-formula characterising regularity can only do so for graphs of bounded size:

\[\phi^{\Cplain}_{\text{regular}(n)} \coloneqq \bigvee_{0 \le i \le n} \forall x \exists^{=i} y \, E(x,y).\]

However, on graphs of fixed size, every \FOC-formula can be simulated by (a family of) \Cplain-formulae \cite[Proposition 8.4.18]{ebbinghaus2005finite}, which is foundational for our proof of \cref{theo:lreceq_to_c_k}.

In itself, $\Cplain$ is exactly as expressive as $\FO$  considering that
\[\exists^{\ge n} x\, \phi(x) \equiv \exists x_1 \dots \exists x_n\, \left[\bigvee_{\substack{i,j \in [n] \\ i \neq j}} x_i \neq x_j \land \bigwedge_{i \in [n]} \phi(x_i) \right].\]
However, we are interested in the finite-variable fragments of $\Cplain$,
denoted as $\Ck{}$ for $k \in \N$, where $\Ck{}$ contains exactly those formulae from $\Cplain$ which only use variables from $\set{x_1,\dots,x_k}$.

We inductively define the \emph{quantifier depth} $\qd(\phi)$ of a formula $\phi \in \Cplain$ as

\[
    \qd(\phi) \coloneqq \begin{cases}
        0 & \text{if } \phi \text{ is atomic}, \\
        \qd(\psi) & \text{if } \phi = \neg \psi, \\
        \max(\qd(\psi_1),\qd(\psi_2)) & \text{if } \phi = \psi_1 \lor \psi_2, \\
        \qd(\phi) + 1 & \text{if } \phi = \exists^{(\ge n)}x \, \psi.
    \end{cases}
\]

Now, we can further restrict $\C{k}{}$ to $\C{k}{r}$, which we define as the subset of formulae $\phi$ of $\C{k}{}$ with $\qd(\phi)\le r$.
\begin{example} The $\C{2}{3}$-sentence
    \[\exists x \, \exists^{=3} y \, \left[ E(x,y) \land \exists^{=4} x \, E(y,x)\right]\]
    expresses that a graph satisfying it must admit a vertex with exactly three neighbours, each in turn admitting exactly 4 neighbours.
\end{example}
In the following, we will mostly be using the asymptotic notation $\C{k}{\bigO{\log n}}$, which should be understood as follows.
Let $\tau$ be a vocabulary, and $\mathfrak{T}$ the class of all $\tau$-structures.
For any $m \in \N$ and $\tau$-structure $\mathcal{B} \in \mathfrak{T}$ we denote, for all $\bar{b} \in V(\mathcal{B})^m$, the pair of a structure and an $m$-tuple of its elements as $\left(\mathcal{B},\bar{b}\right)$.
Let $\mathfrak{B}$ be a class containing a subset of those pairs and suppose $\mathfrak{B}$ to be containing exactly those elements having some property $P$.
We then say that $P$ \emph{can be expressed in $\logCk$} if there exists a $k \in \N$
and a function $f(n) \in \bigO{\log n}$ such that for all $n \in \N$,
there exists a formula $\phi^{(n)}_{\mathfrak{B}}(\bar{v}) \in \C{k}{f(n)}[\tau]$
satisfying for all $(\mathcal{B},\bar{b})$ with $|\mathcal{B}| = n$ that
\(\mathcal{B} \models \phi^{(n)}_{\mathfrak{B}}(\bar{b}) \iff \left(\mathcal{B}, \bar{b}\right) \in \mathfrak{B}.\)

\subsection{The Weisfeiler--Leman Algorithm}

The \emph{($k$-dimensional) Weisfeiler--Leman Algorithm} ($k$\emph{-WL}, $\WL{k}{}$) is a combinatorial algorithm
that iteratively computes a colouring $c^*_k\colon V(G)^k \to C$ on the $k$-vertex tuples of a graph.
Applied to two graphs, it may be used to decide whether these are isomorphic or not.
For our purposes, it suffices to know that the algorithm is initialised 
by colouring all $k$-tuples of vertices by their \emph{atomic type}, which contains all information regarding connectivity and equality of the elements of such a tuple.
The colouring is then iteratively \emph{refined} by computing, for each $k$-tuple of vertices, a new colour based on the colours of the adjacent (that is, differing in one position) $k$-tuples.
The final output is the first colouring $c_k^i$ that partitions the $k$-tuples into the same colour classes as the previous iteration.
More details can be found, for example, in \cite{grohe2021color,kiefer:phd}.

We denote by $\WL{k}{r}$ the restriction of $\WL{k}{}$ that terminates after the first $r$ refinement rounds, \ie, with $c^r_k$;
$\WL{k}{r}$ then \emph{distinguishes} two graphs $G$, $H$ if there exists a colour $c$ such that after $r$ rounds, $G$ and $H$ admit a different number of vertex $k$-tuples of that colour. $\WL{k}{r}$ then \emph{identifies} $G$ if it distinguishes it from all non-isomorphic graphs $H$. The connection between the Weisfeiler--Leman algorithm and the finite-variable fragment of counting logic is as follows.

\begin{lemma}[\cite{grohe2006pebbles,cai1992optimal}]\label{lemm:equivalence_logwl_cklog}
    Let $k \in \N$. For graphs $G$ and $H$ of the same order and all $r \in \N$, the following statements are equivalent:
    \begin{itemize}
        \item $\WL{k}{r}$ distinguishes $G$ and $H$.
        \item There is a $\C{k+1}{r}$-sentence $\phi$ such that $G \models \varphi \iff H \not \models \varphi.$
    \end{itemize}
\end{lemma}

For all $k\in \N$, we define the \emph{logarithmic Weisfeiler--Leman algorithm}, denoted as $\WL{k}{\bigO{\log n}}$, analogous to the asymptotic notation in $\logCk$.
In particular, \cref{lemm:equivalence_logwl_cklog} implies that if there is a $k \ge 2$ and a $\logCk$-sentence identifying a graph $G$,
then $\WL{k-1}{\bigO{\log n}}$ identifies $G$.
 
\section{The Treelike Decompositions}
\label{sec:decomposition}
This section serves us to compute treelike decompositions of logarithmic height, along which we build our \(\Cplain\)-formulae of logarithmic quantifier depth in Section \ref{sec:lrec-to-ck}. We start off with a DAG since, as we will see in Section \ref{sec:lrec-to-ck}, the computation of the relation \(X\) from the definition of \LREC results in such a graph. To account for \(X\) in \(\Cplain\), we can decompose the graph as we are about to describe next. Crucially, the trees underlying the decomposition have logarithmic depth. We will then use the decomposition to construct $\logCk$-formulae.

Throughout this section,
we assume that \(G\) is a rooted DAG and that \(r\) is the root of \(G\).

The \emph{tree unfolding} of \(G\) is the tree \(T_G\) whose vertices are paths
\(\bar{v} = (v_0, \dots, v_k)\) in \(G\) with \(v_0=r\),
and where \(\bar{w} = (w_0, \dots, w_\ell)\) is a child of \(\bar{v}\)
if \(\ell = k+1\) and \(w_i = v_i\) for $i\in[0,k]$.
For every \(v \in V(G)\), let \(\CP_G(v)\) be the set of all paths
\((v_0, \dots, v_k)\) in \(G\) with \(v_0 = r\) and \(v_k = v\).
Note that all \(\bar{v} \in \CP(v)\) are vertices of \(T_G\).
We call them the \emph{copies} of \(v\) in \(T_G\).

We define the \emph{weight} of a vertex \(v\) in \(G\) to be
\[
  \wt_G(v) \deff \abs{\CP(v)},
\]
of which we omit the subscript if it is clear from context.
We define the \emph{aggregate weight} of \(G\) to be
\(\awt(G) \deff \sum_{v \in V(G)} \wt(v)\).
Observe that
\[
  \awt(G)=\abs{T_G}.
\]
We define the \emph{multiplicity} of a vertex \(v\) to be
\[
    \mul_G(v) \deff \max \left\{\left.
    \prod_{i=1}^k\deg^-(v_i)\;\right|\;
    (v_0, \dots,v_k)\in \CP(v)\right\},
\]
where the empty product is \(1\) (thus \(\mul_G(r) = 1\)). We may again omit the subscript.
Further, we let the \emph{aggregate multiplicity} of $G$ be defined as \(\amul(G) \deff \sum_{v \in V(G)} \mul(v)\).

\begin{lemma}\label{lem:1}
  For all \(v \in V(G)\), we have
  $
    \wt(v) \leq \mul(v).
  $
\end{lemma}

\begin{IEEEproof}
  We prove the result by induction on the distance between \(v\) and the root \(r\).
  For the root \(r\), we have \(\wt(r) = \mul(r) = 1\).
  So let \(v\) be a node with in-neighbours \(u_1, \dots, u_k\).
  Then \(\deg^-(v) = k\) and
  \[
    \wt(v) = \sum_{i=1}^k \wt(u_i)
           \leq \sum_{i=1}^k \mul(u_i) 
           \leq k \max_{i \in [k]} \mul(u_i)
           = \mul(v),
  \]
  where the first inequality holds by the induction hypothesis.
\end{IEEEproof}

For \(m \in \N\), we say that \(G\) has the \emph{\(m\)-path property}
if \(\mul(v) \leq m\) for all \(v \in V(G)\). The $m$-path property allows us to control the size of a tree unfolding.

\begin{corollary}
  \label{cor:m-path-property}
  If \(G\) has the \(m\)-path property,
  then \(\awt(G) \leq \amul(G) \leq m \cdot \abs{G}\).
\end{corollary}

Let \(v \in V(G)\) and \(W \subseteq V(G)\) be such that
\(v \dagle_G w\) for all \(w \in W\).
Then, we define \(G_v^W\) to be the induced subgraph of \(G\) with vertex set
\begin{align*}
  \bigl\{u \in V(G) \bigmid &\text{ there is path } (v_0, \dots, v_\ell)
    \text{ in } G \text{ with } v_0 = v \text{ and}\\
    &\ v_\ell = u  \text{ and } \set{v_0, v_1, \dots, v_{\ell-1}} \cap W =\varnothing
  \big\}.
\end{align*}

In the definition, we stipulate that for \(\ell = 0\),
\(\set{v_0, v_1, \dots, v_{\ell-1}}\) is the empty set.
Thus, if \(v \in W\), then \(G_v^W = G[\set{v}]\).
Note that \(G_v^W\) is a rooted DAG with root \(v\).
If \(W = \varnothing\), we write \(G_v\) instead of \(G_v^W\),
and if \(v\) is the root, we write \(G^W\) instead of \(G_r^W\).
If \(W = \set{w_1, \dots, w_k}\),
we also write \(G_v^{w_1, \dots, w_k}\) instead of \(G_v^W\).

\begin{lemma}\label{lem:2}
  For all \(v \in V(G)\), it holds that \(\awt(G_v) + \awt(G^v) \leq \awt(G)+1\).
\end{lemma}

\begin{IEEEproof}
  Let \(X \deff V(G_v) \setminus V(G^v)\),
  \(Y \deff V(G^v) \setminus V(G_v)\),
  and \(Z \deff \big(V(G_v) \cap V(G^v)\big) \setminus \set{v}\). 
  We have
  \begin{align*}
    \awt(G)   &&= &&\wt_G(v)
              &&+ &&\sum\limits_{x \in X} \wt_G(x)\\
              &&+ &&\sum\limits_{y \in Y} \wt_G(y)
              &&+ &&\sum\limits_{z \in Z} \wt_G(z)\\
    \awt(G_v) &&= &&\wt_{G_v}(v)
              &&+ &&\sum\limits_{x \in X} \wt_{G_v}(x)\\
              &&  &&
              &&+ &&\sum\limits_{z \in Z} \wt_{G_v}(z)\\
    \awt(G^v) &&= &&\wt_{G^v}(v)
              &&  &&\\
              &&+ &&\sum\limits_{y \in Y} \wt_{G^v}(y)
              &&+ &&\sum\limits_{z \in Z} \wt_{G^v}(z).
  \end{align*}
  We have \(\wt_{G_v}(v) = 1\) and \(\wt_{G^v}(v) = \wt_G(v)\).
  Furthermore, for every \(z \in Z\),
  we have \(\wt_{G_v}(z) + \wt_{G^v}(z) \leq \wt_G(z)\),
  because we can partition \(\CP_G(z)\) into \(\CP_{G^v}(z)\),
  consisting of all paths from \(r\) to \(z\) in \(G\) that avoid \(v\),
  and the set \(\CQ\) consisting of all paths
  from \(r\) to \(z\) in \(G\) that contain \(v\).
  We have \(\abs{\CP_{G_v}(z)} \leq \abs{\CQ}\).
  With this, the assertion of the lemma follows.
\end{IEEEproof}

\begin{lemma}\label{lem:3}
  Let \(v \in V(G)\) such that \(v\) is not a leaf of \(G\).
  Then there is an \(a \in V(G)\) such that \(v \dagle a\) and
  \begin{align}
    \label{eq:8}
    \awt(G_v^a) &\leq \frac{\awt(G_v)}{2},\\
    \label{eq:9}
    \awt(G_b) &\leq \ceil{\frac{\awt(G_v)}{2}} &\text{for all \(b \in N^+(a)\)}.
  \end{align}
\end{lemma}

\begin{IEEEproof}
  Let \(m \deff \awt(G_v)\) and note that \(m \geq \abs{G_v} \geq 2\).

  Then \(\awt(G_v^v) = 1 \leq \frac{m}{2}\)
  and \(\awt(G_v^w) = \awt(G_v) = m > \frac{m}{2}\)
  for every leaf \(w\) of \(G_v\).
  Hence, there is an \(a \in V(G_v)\) such that
  \(\awt(G_v^a) \leq \frac{m}{2}\)
  and \(\awt(G_v^b) > \frac{m}{2}\) for every \(b \in N^+(a)\).

  This \(a\) satisfies \eqref{eq:8};
  to see that it satisfies \eqref{eq:9}, let \(b \in N^+(a)\).
  Then \(\awt(G_v^b) > \frac{m}{2}\) and
  thus \(\awt(G_v^b) \geq \floor{\frac{m}{2}}+1\).
  By \cref{lem:2}, we have \(\awt(G_b) + \awt(G_v^b) \leq m+1\)
  and thus
  \[
    \awt(G_b) \leq m+1 - \left(\floor*{\frac{m}{2}}+1\right)
    = \ceil{\frac{m}{2}}.
  \]
\end{IEEEproof}

\begin{lemma}\label{lem:4}
  Let \(v, w \in V(G)\) be such that \(v \dagsle w\).
  Then there is an \(a \in V(G)\) such that \(v \dagle a \dagsle w\) and
  \begin{align}
    \label{eq:6}
    \awt(G_v^a) &\leq \frac{\awt(G^w_v)}{2},\\
    \label{eq:7}
    \awt(G_b^w) &\leq \ceil{\frac{\awt(G^w_v)}{2}}
                &\hspace{-1.7ex}\text{for all \(b \in N^+\mspace{-3mu}(a)\) with \(b\dagle w\)}.
  \end{align}
\end{lemma}

\begin{IEEEproof}
  Let \(m \deff \awt(G^w_v)\) and note that \(m \geq \abs{G^w_v} \geq 2\).
  Let \(\bar{v} = (v_0, \dots, v_k)\)
  be a path in \(G\) with \(v_0 = v\) and \(v_k = w\).
  We have \(\awt(G_v^{v_0}) = 1 \leq \frac{m}{2}\)
  and \(\awt(G_v^{v_k}) = \awt(G^w_v) > \frac{m}{2}\).
  Thus there is a (unique) \(i \in \set{0, \dots, k-1}\) such that
  \(\awt(G_v^{v_i}) \leq \frac{m}{2}\)
  and \(\awt(G_v^{v_{i+1}}) > \frac{m}{2}\).
  Let \(a(\bar{v}) \deff v_i\),
  and let \(a\) be \(\dagle\)-maximal among all \(a(\bar{v})\),
  where \(\bar{v}\) ranges over all paths from \(v\) to \(w\).

  Then \eqref{eq:6} is trivially satisfied by all \(a(\bar{v})\)
  and in particular by \(a\).

  To prove \eqref{eq:7}, let \(b \in N^+(a)\) such that \(b \dagle w\).
  As \(v \dagle a\) and \(ab \in E(G)\) and \(b \dagle w\),
  there exists a path \((v_0, \dots, v_k)\) from \(v\) to \(w\) such that
  \(a = v_i\) and \(b = v_{i+1}\).
  By the maximality of \(a\), we have \(\awt(G_v^b) > \frac{m}{2}\).
  By \cref{lem:2} applied to the graph \(G_v^w\) and \(b\),
  we get \(\awt(G_b^w) \leq \ceil{\frac{m}{2}}\).
\end{IEEEproof}

We can now use \cref{lem:3,lem:4} to inductively construct a representation of \(G\) by a tree of logarithmic height.

\begin{lemma}\label{lem:5}
  There are a rooted tree \(T\) and mappings
  \(v \colon V(T) \to V(G)\), \(W \colon V(T) \to 2^{V(G)}\)
  such that the following conditions are satisfied.
  \begin{enumerate}
    \item
      \label{item:at-most-one}
      \(|W(t)| \leq 1\) for all \(t \in V(T)\).
    \item
      \label{item:leaf}
      \(t \in V(T)\) is a leaf of \(T\) if and only if
      \(v(t)\) is a leaf of \(G\) or \(W(t) = \set{v(t)}\).
    \item
      \label{item:no-leaf}
      If \(t \in V(T)\) is not a leaf of \(T\) and \(W(t) = \set{w}\),
      then \(v(t) \dagsle w\).
    \item
      \label{item:decomposition}
      If \(t \in V(T)\) with children \(u_1, \dots, u_k\) for some \(k\geq 1\),
      then
      \[
        V\left(G_{v(t)}^{W(t)}\right) \setminus \set{v(t)}
        \subseteq \bigcup_{i=1}^k V\left(G_{v(u_i)}^{W(u_i)}\right).
      \]
    \item
      \label{item:height}
      The height of \(T\) is at most \(2\log\bigl(\awt(G)\bigr)\).
  \end{enumerate}
\end{lemma}

\begin{IEEEproof}
  We define the tree \(T\) inductively.
  We start with a root \(r_T\) and let
  \(v(r_T) \deff r\) and \(W(r_T) \deff \varnothing\).

  To extend the tree,
  let \(t\) be a node in \(T\) where the children are not yet defined.
  If \(v(t)\) is a leaf of \(G\) or \(v(t) \in W(t)\),
  then \(t\) is a leaf of \(T\).
  Now suppose that \(v(t)\) is not a leaf of \(G\) and \(v(t) \not \in W(t)\).
  Let \(v \deff v(t)\) and \(W \deff W(t)\).
  By induction, we assume
  \(\abs{W} \leq 1\) and \(v \dagsle_G w\) if \(W=\set{w}\).
  \begin{description}[\IEEEsetlabelwidth{Case 1:}]
    \item[\emph{Case 1:}]
      \(W = \varnothing\).
      (We say that \(t\) is a node of type \(0\).)\\
      Then \(G_v^W = G_v\).
      By \cref{lem:3}, there is an \(a \in V(G_v)\) such that
      \(\awt(G_v^a) \leq \frac{\awt(G_v)}{2}\)
      and \(\awt(G_b) \leq \ceil{\frac{\awt(G_v)}{2}}\)
      for all \(b \in N^+(a)\).

      We add a child \(u_a\) of \(t\) with \(v(u_a) \deff v\) and
      \(W(u_a) \deff \set{a}\).
      For every \(b \in N^+(a)\), we add a child \(u_b\)
      with \(v(u_b) \deff b\) and \(W(u_b) \deff \varnothing\).
    \item[\emph{Case 2:}]
      \(W =\set{w}\) for some \(w\).
      (We say that \(t\) is a node of type \(1\).)\\
      Then  \(G_v^W = G_v^w\).
      By \cref{lem:4}, there is an \(a \in V(G_v)\) such that
      \(\awt(G_v^a) \leq \frac{\awt(G^w_b)}{2}\)
      and \(\awt(G_b) \leq \ceil{\frac{\awt(G^w_v)}{2}}\)
      for all \(b \in N^+(a)\) with \(b \dagle w\).

      We add a child \(u_a\) of \(t\) with \(v(u_a) \deff v\) and
      \(W(u_a) \deff \set{a}\).
      For every \(b \in N^+(a)\) with \(b \dagle w\),
      we add a child \(u_b\) with \(v(u_b) \deff b\)
      and \(W(u_b) \deff \set{w}\).
      For every \(b \in N^+(a)\) with \(b \not\dagle w\),
      we add a child \(u_b\) with \(v(u_b) \deff b\)
      and \(W(u_b) \deff \varnothing\).
  \end{description}

  It is immediate from the construction that \(T, v, W\) satisfy
  \cref{item:at-most-one,item:leaf,item:no-leaf,item:decomposition}
  of \cref{lem:5}.
  We need to prove that they satisfy \cref{item:height}.
  For every \(t \in V(T)\), let \(A(t) \deff \awt\left(G_{v(t)}^{W(t)}\right)\).
  Observe that for all nodes \(t \in V(T)\)
  and all children \(u\) of \(t\) the following holds:
  \begin{itemize}
    \item \(A(u) < A(t)\);
    \item if \(t\) is of type \(0\),
      then \(A(u) \leq \ceil{\frac{A(t)}{2}}\);
    \item if \(A(u) > \ceil{\frac{A(t)}{2}}\),
      then \(t\) is of type \(1\) and \(W(u) = \varnothing\),
      so \(u\) is of type \(0\).
  \end{itemize}
  This implies that for all grandchildren \(v\) of \(t\),
  it holds that \(A(v) \leq \frac{A(t)}{2}\),
  and as \(A(r_T) = \awt(G)\),
  \cref{item:height} follows.
\end{IEEEproof}
 
\section{From \LRECeq to \logCk}
\label{sec:lrec-to-ck}
Let \(G\) be a directed graph.
A \emph{cardinality condition}
for \(G\) is a mapping \(C\)
that associates to each \(v \in V(G)\)
a set \(C(v) \subseteq [0, \deg^+(v)]\).
Given a cardinality condition \(C\),
we define, analogously to the definition of \LRECeq, \(X = X(G,C) \subseteq V(G) \times \Npos\)
to be the inclusionwise smallest set such that
for all \(v \in V(G)\) and \(i \in \Npos\),
it holds that \((v, i) \in X\) if and only if
\[
  \left|\left\{w \in N^+(v)\;\left|\;
    \left(w,\floor*{\frac{i-1}{\deg^-(w)}}\right) \in X
  \right.\right\}\right| \in C(v).
\]

For every \(n \in \Npos\), we define a vocabulary
\(\tau^{(n)} \deff \set{E, P_0, \dots, P_n}\),
where \(E\) is a binary relation symbol
and the \(P_i\) are unary relation symbols.
We can represent a tuple \((G,C)\) consisting of
a directed graph \(G\) of order \(\abs{G} \leq n\)
and a cardinality condition \(C\) for \(G\)
as a \(\tau^{(n)}\)-structure \(\A = \A(G,C)\)
with \(V(\A) \deff V(G)\), \(E(\A) \deff E(G)\),
and, for all \(i \in [0,n]\),
\[P_i(\A) \deff \setc{v \in V(G)}{i \in C(v)}.\]

The following theorem enables us to check $X$-membership via formulae in counting logics with logarithmic quantifier depth.

\begin{theorem}\label{theo:Ck-for-X}
  There is a \(k \in \Npos\) such that for all \(n, r \in \Npos\)
  and \(i \in [(n+1)^r]\),
  there is a \(\Ck{\bigO{r \log n}}\)-formula \(\phi^{(n)}_i(x)\)
  such that for all directed graphs \(G\) of order \(\abs{G} \leq n\),
  all cardinality conditions \(C\) for \(G\),
  and all \(v \in V(G)\),
  it holds that
  \[
    \A(G,C) \models \phi^{(n)}_i(v) \iff (v,i) \in X(G,C).
  \]
\end{theorem}
\begin{figure}
    \centering
    \begin{tikzpicture}
            \node[] (C) at (-2,10) {$H_{v,i}$};
            
            \node[draw=black, rounded corners] (1) at (0,10) {$(v,i)$};
            
            \node[draw=black, rounded corners] (2) at (-2,8) {$\left(w_1,\floor*{\frac{i - 1}{\deg_G^-(w_1)}}\right)$};
            \node[] (3) at (0,8) {$\dots$};
            \node[draw=black, rounded corners] (4) at (2,8) {$\left(w_m,\floor*{\frac{i - 1}{\deg_G^-(w_m)}}\right)$};
            
            \node[] (5) at (-2,7) {$\vdots$};
            \node[] (6) at (2,7) {$\vdots$};

            \draw[->] (1) -- (2);
            \draw[->] (1) -- (4);
    \end{tikzpicture}
    \caption{The recursive construction of $H_{v,i}$: in this example, it holds that $N^+_G(v) = \set{w_1,\dots,w_m}$.}
    \label{fig:lrec_tree}
\end{figure}
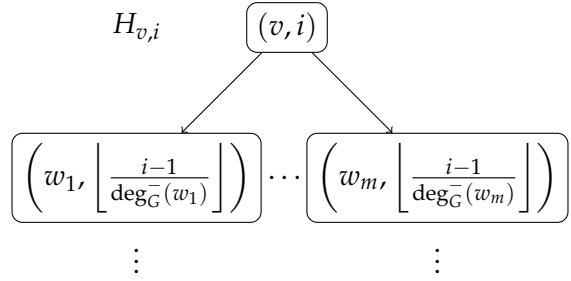 
\begin{IEEEproof}
  Let \(n, r \in \Npos\), and let \(i \in [(n+1)^r]\).
  First, for every directed graph \(G\) of order \(\abs{G} \leq n\)
  and every cardinality condition \(C\) on \(G\),
  we are going to describe, for all \(v \in V(G)\), rooted DAGs \(H_{v, i}\)
  that may be used to decide whether \((v, i)\) is contained in \(X(G,C)\).
  Then, \cref{lem:5} yields trees \(T_{v, i}\)
  of logarithmic height based on the \(H_{v, i}\).
  We describe how to use those trees to decide whether \((v, i) \in X(G,C)\) holds. 
  At the end of this proof,
  we recursively construct formulae that check containment in \(X(G,C)\) and have a structure that closely follows the structure
  of the described trees.
  Since the tree from \cref{lem:5} has logarithmic height,
  the formulae will have a logarithmic quantifier depth.

  Let \(G\) be a directed graph of order
  \(\abs{G} \leq n\), and let \(v \in V(G)\).
  We inductively define a rooted DAG \(H_{v, i}\),
  see also \cref{fig:lrec_tree}.
  We start with the root \((v, i)\).
  Then, repeatedly, for every vertex \((v', i') \in V(H_{v, i})\)
  and every neighbour \(w \in N_G^+(v')\)
  where for \(j \deff \floor*{\frac{i' - 1}{\deg_G^-(w)}}\), it holds that \(j \geq 1\),
  we add a vertex \((w, j)\)
  to \(H_{v, i}\) (unless it already exists)
  and insert an edge from \((v', i')\) to \((w, j)\).

  We could decide ``\((v, i) \in X(G,C)\)?'' as follows. 
  First, we go through all leaves \((w, j)\) in \(H_{v, i}\)
  and mark them as positive if \(0 \in C(w)\) and as negative otherwise.
  Then, for every vertex \((w, j)\) that has only marked children,
  we mark the vertex as positive if and only if the number of positively marked
  children is contained in \(C(w)\), and we mark it as negative otherwise.
  Once all vertices have been marked, we have \((v, i) \in X(G,C)\)
  if and only if \((v, i)\) is marked as positive.
  Since the height of \(H_{v, i}\) may be linear in the size of \(G\),
  this process might take a linear number of steps.
  Thus, we use a tree \(T_{v, i}\) of logarithmic height instead,
  which we describe below.
  
  Note that for a node \((w, j) \in V(H_{v, i})\), the graph 
  \(H_{w, j}\) is the induced subgraph of \(H_{v, i}\)
  on all nodes below (or equal to) \((w, j)\).
  
  \begin{claim}\label{clm:path-property}
    \(H_{v, i}\) has the \((n+1)^r\)-path property.
  \end{claim}

  \begin{claimproof}
    We prove the equivalent statement
    \(\mul_{H_{v, i}}\bigl((w, j)\bigr) \leq \bigl(\abs{H_{v, i}}+1\bigr)^r\)
    for all \((w, j) \in V(H_{v, i})\).
    
    Let \((v_0', \dots, v_p')\) be a path in \(H_{v, i}\) with
    \(v_0' = (v, i)\) and \(v_p' = (w, j)\).
    Moreover, let \((v_s, \ell_s) \deff v_s'\) for all \(s \in [0,p]\).
    Then, we have
    \[\ell_s = \floor*{\frac{\ell_{s-1} - 1}{\deg_G^-(v_s)}}
    \leq \frac{\ell_{s-1}}{\deg_G^-(v_s)} = \frac{i}{\prod_{t=1}^s \deg_G^-(v_t)}\]
    for all \(s \in [p]\).
    With \(\ell_p = j \geq 1\), it holds that
    \(i \geq \prod_{s=1}^p \deg_G^-(v_s)
    \geq \prod_{s=1}^p \deg_{H_{v, i}}^-(v_s)\).
    Thus, \(\mul_{H_{v, i}}\bigl((w, j)\bigr) \leq i \leq (n+1)^r\).
    Hence, \(H_{v, i}\) has the \((n+1)^r\)-path property.
  \end{claimproof}

  \begin{figure*}
    \centering
    \begin{subfigure}[b]{.3\textwidth}
        \centering
        \begin{tikzpicture}
            \node[] (C) at (-1.75,10) {$H_{v,i}$};
            \node[draw=black, rounded corners] (1) at (0,10) {$(v,i)$};
            \node[draw=black, rounded corners] (2) at (0,8.5) {$(v',i')$};
            \node[draw=black, rounded corners] (3) at (0,7) {$\textcolor{marine}{(a,\ell)}$};

            \node[draw=black, rounded corners] (4) at (-1.5,6) {$\textcolor{red}{(b_1,\ell'_1)}$};
            \node[] (5) at (0,6) {$\dots$};
            \node[draw=black, rounded corners] (6) at (1.5,6) {$\textcolor{green}{(b_m,\ell'_m)}$};
            
            \draw[decorate,decoration={snake},->] (1) -- (2);
            \draw[decorate,decoration={snake},->] (2) -- (3);
            \draw[->] (3) -- (4);
            \draw[->] (3) -- (6);
            \draw[decorate,decoration={snake},->,opacity=0.2]    (2) to[out=-160,in=90] (4);
            \draw[decorate,decoration={snake},->,opacity=0.2]    (2) to[out=-20,in=90] (6);
        \end{tikzpicture}
    \end{subfigure}
    \begin{subfigure}[b]{.0\textwidth}
        \centering
        \begin{tikzpicture}
            \node[] (1) at (0,5) {};
            \node[] (2) at (2,5) {};
            \node[] (3) at (0,5) {};
            \node[] (4) at (0,1) {};

            \draw[->,ultra thick]    (1) to[out=20,in=160] (2);
        \end{tikzpicture}
    \end{subfigure}
    \begin{subfigure}[b]{.68\textwidth}
        \centering
        \begin{tikzpicture}
            \node[] (C) at (-2,10) {$T_{v,i}$};
            \node[] (1) at (0,10) {$r:((v,i),\varnothing)$};
            \node[draw=black] (C) at (3,10) {$t:(v(t),W(t))$};
            \node[] (2) at (0,8.5) {$t:((v',i'),\varnothing)$};
            
            \node[] (3) at (-4,7) {$u_0: ((v',i'),\set{\textcolor{marine}{(a,\ell)}})$};
            \node[] (4) at (-0.5,7) {$u_1: (\textcolor{red}{(b_1,\ell'_1)},\varnothing)$};
            \node[] (5) at (1.25,7) {$\dots$};
            \node[] (6) at (3,7) {$u_m: (\textcolor{green}{(b_m,\ell'_m)},\varnothing)$};
            \node[] (7) at (3,6.25) {};
            
            \draw[decorate,decoration={snake},->] (1) -- (2);
            \draw[->] (2) -- (3);
            \draw[->] (2) -- (4);
            \draw[->] (2) -- (6);
        \end{tikzpicture}
    \end{subfigure}
    \caption{Letting $(v,i)$ be the vertex for which we want to know whether ``$(v,i) \in X(G,C)$?'', the figure shows how to recursively obtain $T_{v,i}$ at a vertex $t \in V(T_{v,i})$ of type $0$ with $v(t) = (v',i'),W = \varnothing$.}
    \label{fig:type0}
\end{figure*}
 
  Next, we apply \cref{lem:5} to \(H_{v, i}\) and obtain (the existence of)
  a rooted tree \(T_{v, i}\) and mappings \(v \colon V(T_{v, i}) \to V(H_{v, i})\) and
  \(W \colon V(T_{v, i}) \to 2^{V(H_{v, i})}\).
  Let \(T_{v, i}, v, W\) be as described in the proof of \cref{lem:5}.
  
  Now, we describe how to decide ``\((v, i) \in X(G,C)\)?'' using \(T_{v, i}\).
  Let \(H \deff H_{v, i}\) and \(T \deff T_{v, i}\).
  We start with the root \(r\) of \(T\), which, by the construction from \cref{lem:5},
  is a node of type \(0\) (that is, at node \(t\) with \(W(t) = \varnothing\)) with \(v(r) = (v, i)\) and \(W(r) = \varnothing\). At every node \(t \in V(T)\) of type \(0\) with \(v(t) = (v', i')\) and \(W(t) = \varnothing\),
  our goal is to decide whether \((v', i') \in X(G,C)\) by recursively checking the children of \(t\).
  At every node \(t \in V(T)\) of type \(1\) with \(v(t) = (v', i')\) and \(W(t) = \set{(w, j)}\),
  we are additionally given a number \(c\)
  and our goal is to decide whether \((v', i') \in X(G,C)\) by recursively checking the children of \(t\)
  under the assumption that exactly \(c\) of the children of \((w, j)\) in \(H\) are contained in \(X(G,C)\).
  In detail, the computations work as follows.
  
  For the following, see also \cref{fig:type0}. Consider a node \(t \in V(T)\) of type \(0\) with children
  \(u_0\) of type \(1\) and \(u_1, \dots, u_m\) of type \(0\).
  Let \(v(t) = (v', i')\), \(W(t) = \varnothing\),
  \(v(u_0) = (v', i')\), \(W(u_0) = \set{(a, \ell)}\),
  and \(v(u_s) =(b_s, \ell'_s)\), \(W(u_s) = \varnothing\) for all \(s \in [m]\).
  By the construction of \(T\) in \cref{lem:5},
  we have \((v,i) \dagle_H (v', i') \dagsle_H (a, \ell) \dagsle_H (b_s, \ell'_s)\) for all \(s \in [m]\),
  where the \((b_s, \ell'_s)\) are children of \((a, \ell)\) in \(H\).

  To decide whether ``\((v', i') \in X(G,C)\)?'',
  we first decide \((b_s, \ell'_s) \in X(G,C)\) recursively for all \(s \in [m]\).
  This is equivalent to running our procedure recursively on the children \(u_1, \dots, u_m\) of \(t\) of type \(0\).
  Let \(c\) be the number of the \((b_s, \ell'_s)\) contained in \(X(G,C)\).
  We then run our procedure recursively on the child \(u_0\) of type \(1\)
  with \(v(u_0) = (v', i')\) and \(W(u_0) = \set{(a, \ell)}\)
  to check whether \((v', i')\) is contained in \(X(G,C)\)
  under the assumption that exactly \(c\) of the children of \((a, \ell)\)
  are contained in \(X(G,C)\).

  Now consider a node \(t \in V(T)\) of type \(1\) with children
  \(u_0\) of type \(1\), \(u_1, \dots, u_m\) of type \(1\),
  and \(u_{m+1}, \dots, u_{m+p}\) of type \(0\) (see \cref{fig:type1}).
  Let \(v(t) = (v', i')\), \(W(t) = \set{(w, j)}\),
  \(v(u_0) = (v', i')\), \(W(u_0) = (a, \ell)\),
  \(v(u_s) =(b_s, \ell'_s)\), \(W(u_s) = \set{(w, j)}\) for all \(s \in [m]\),
  and \(v(u_s) =(b_s, \ell'_s)\), \(W(u_s) = \varnothing\) for all \(s \in [m+1,p]\).
  By the construction of \(T\) in \cref{lem:5},
  we have \((v,i) \dagle_H (v', i') \dagsle_H (a, \ell) \dagsle_H (b_s, \ell'_s)\) for all \(s \in [m+p]\),
  where the \((b_s, \ell'_s)\) are children of \((a, \ell)\) in \(H\);
  we have \((b_s, \ell'_s) \dagle_H (w, j)\) for all \(s \in [m]\) and
  \((b_s, \ell'_s) \not\dagle_H (w, j)\) for all \(s \in [m+1, p]\).

  To decide whether ``\((v', i') \in X(G,C)\)?'' under the assumption
  that exactly \(c\) children of \((w, j)\) in \(H\) are contained in \(X(G,C)\),
  we recursively run our procedure on the children \(u_1, \dots, u_m\) of type \(1\)
  under the above-mentioned assumption and thereby decide
  \((b_s, \ell'_s) \in X(G,C)\) for \(s \in [m]\).
  Next, we recursively run the procedure on the children \(u_{m+1}, \dots, u_{m+p}\) of type \(0\)
  and thereby decide \((b_s, \ell'_s) \in X(G,C)\) for \(s \in [m+1, m+p]\).
  Let \(c'\) be the number of \((b_s, \ell'_s) \in X(G,C)\) for \(s \in [m+p]\).
  Then, finally, we run our procedure recursively on the child \(u_0\) of type \(1\)
  to check whether \((v', i')\) is contained in \(X(G,C)\)
  under the assumption that exactly \(c'\) of the children of \((a, \ell)\)
  are contained in \(X(G,C)\).

  For now, we assumed that the considered nodes are not leaves.
  For a leaf \(t\) of type \(0\) with \(v(t) = (v', i')\) and \(W(t) = \varnothing\),
  we have \((v', i') \in X(G,C)\) if and only if \(0 \in C(v')\)
  since \((v', i')\) is a leaf of \(H_{v, i}\) by \cref{lem:5} (\cref{item:leaf}).
  For a leaf \(t\) of type \(1\) with \(v(t) = (v', i')\),
  by \cref{lem:5} (\cref{item:leaf}),
  we have \(W(t) = \set{(v', i')}\).
  Hence, we have to decide \((v', i') \in X(G,C)\) under the assumption that
  exactly \(c\) children of \((v', i')\) in \(H\) are contained in \(X(G,C)\)
  for some number \(c\).
  This holds if and only if \(c \in C(v')\).

  Before we translate the structure of the trees \(T_{v, i}\) into formulae,
  we first check that the trees are of logarithmic height.
  This is essential to obtain formulae of logarithmic quantifier depth.

  \begin{figure*}
    \centering
    \begin{subfigure}[b]{.34\textwidth}
        \centering
        \begin{tikzpicture}
            \node[] (C) at (-2,10) {$H_{v,i}$};
            \node[draw=black, rounded corners] (1) at (0,10) {$(v,i)$};
            \node[draw=black, rounded corners] (2) at (0,8.5) {$(v',i')$};
            \node[draw=black, rounded corners] (3) at (0,7) {$\textcolor{marine}{(a,\ell)}$};

            \node[draw=black, rounded corners] (4) at (-3,5.5) {$\textcolor{red}{(b_1,\ell'_1)}$};
            \node[] (5) at (-2,5) {$\ddots$};
            \node[draw=black, rounded corners] (6) at (-1,4.5) {$\textcolor{green}{(b_m,\ell'_m)}$};

            \node[draw=black, rounded corners] (7) at (1,3.25) {$\textcolor{purple}{(b_{m+1},\ell'_{m+1})}$};
            \node[] (8) at (2,4) {\reflectbox{$\ddots$}};
            \node[draw=black, rounded corners] (9) at (3,4.5) {$\textcolor{orange}{(b_{m+p},\ell'_{m+p})}$};

            \node[draw=black, rounded corners] (10) at (-2,3) {$\textcolor{blue}{(w,j)}$};

            \draw[decorate,decoration={snake},->,opacity=0.2]    (2) to[out=-160,in=110] (6);
            \node[fill=white,circle,inner sep=5pt,opacity=0.8] (circ1) at (-1.5,6.25) {};
            \draw[decorate,decoration={snake},->,opacity=0.2]    (2) to[out=-20,in=70] (7);
            \node[fill=white,circle,inner sep=5pt,opacity=0.8] (circ2) at (1.75,5.5) {};
            \draw[decorate,decoration={snake},->,opacity=0.2]    (2) to[out=-170,in=70] (4);
            \draw[decorate,decoration={snake},->,opacity=0.2]    (2) to[out=-10,in=110] (9);
            
            \draw[decorate,decoration={snake},->] (1) -- (2);
            \draw[decorate,decoration={snake},->] (2) -- (3);
            \draw[->] (3) -- (4);
            \draw[->] (3) -- (6);
            \draw[->] (3) -- (7);
            \draw[->] (3) -- (9);

            \draw[decorate,decoration={snake},->] (4) -- (10);
            \draw[decorate,decoration={snake},->] (6) -- (10);
        \end{tikzpicture}
    \end{subfigure}
    \begin{subfigure}[b]{.04\textwidth}
        \centering
        \begin{tikzpicture}
            \node[] (1) at (0,5) {};
            \node[] (2) at (2,5) {};
            \node[] (3) at (0,5) {};
            \node[] (4) at (0,-2) {};

            \draw[->,ultra thick]    (1) to[out=20,in=160] (2);
        \end{tikzpicture}
    \end{subfigure}
    \begin{subfigure}[b]{.6\textwidth}
        \centering
        \begin{tikzpicture}
            \node[] (C) at (-2,10) {$T_{v,i}$};
            \node[draw=black] (C) at (3,10) {$t:(v(t),W(t))$};
            \node[] (1) at (0,10) {$r:((v,i),\varnothing)$};
            \node[] (2) at (0,8.5) {$t:((v',i'),\set{\textcolor{blue}{((w,j)}})$};
            
            \node[] (3) at (-3,7.5) {$u_0:((v',i'),\set{\textcolor{marine}{(a,\ell)}})$};
            \node[] (4) at (-2.5,6) {$u_1:(\textcolor{red}{(b_1,\ell'_1)},\set{\textcolor{blue}{((w,j)}})$};
            \node[] (5) at (-2.25,5) {$\ddots$};
            \node[] (6) at (-2.2,3.3) {$u_m: (\textcolor{green}{(b_m,\ell'_m)},\set{\textcolor{blue}{((w,j)}})$};
            \node[] (7) at (2.2,3.3) {$u_{m+1}:(\textcolor{purple}{(b_{m+1},\ell'_{m+1})},\varnothing)$};
            \node[] (8) at (2.25,5) {\reflectbox{$\ddots$}};
            \node[] (9) at (2.5,6) {$u_{m+p}:(\textcolor{orange}{(b_{m+p},\ell'_{m+p})},\varnothing)$};
            \node[] () at (3,3) {};
            
            \draw[decorate,decoration={snake},->] (1) -- (2);
            \draw[->] (2) -- (3);
            \draw[->] (2) -- (4);
            \draw[->] (2) to[out=-92,in=40] (6);
            \draw[->] (2) to[out=-88,in=140] (7);
            \draw[->] (2) -- (9);
        \end{tikzpicture}
    \end{subfigure}
    \caption{Letting $(v,i)$ be the vertex for which we want to know whether ``$(v,i) \in X(G,C)$?'', the figure shows how to recursively obtain $T_{v,i}$ at a vertex $t \in V(T_{v,i})$ of type $1$ with $v(t) = (v',i'),W = \set{\textcolor{blue}{((w,j)}}$.}
    \label{fig:type1}
\end{figure*} 
  \begin{claim}
    \label{clm:tree-height}
    \(T_{v, i}\) has height at most \((4r+2) \cdot \log (n+1)\).
  \end{claim}

  \begin{claimproof}
    By \cref{clm:path-property},
    \(H_{v, \ell}\) has the \((n+1)^r\)-path property.
    Thus, with \cref{cor:m-path-property},
    we obtain \(\awt(H_{v, \ell}) \leq (n+1)^r \cdot \abs{H_{v, \ell}}\),
    and hence, by \cref{lem:5},
    the height of \(T_{v, \ell}\) is at most
    \(2 \log \bigl((n+1)^r \cdot \abs{H_{v, \ell}}\bigr)
    \leq 2 \log \bigl((n+1)^r \cdot (n+1)^r \cdot n\bigr)
    \leq 2 \cdot (2r+1) \cdot \log (n+1)\).
  \end{claimproof}
  
  In the following, we recursively construct formulae of the form \(\psi^{h,n}_{t0, i'}(x)\)
  and \(\psi^{h,n}_{t1, i', j, c}(x,y)\) with domain variables $x, y$. 
  Before beginning, it is appropriate to shortly analyse the syntax. By $n$, we refer to the order of the structure, which is fixed. With $t0$ resp.\ $t1$, we keep track of whether the formula at hand corresponds to a node of type $0$ or $1$. The number $h \in \N$ tracks our recursion depth and ensures that we do not produce formulae with non-logarithmic depth.

  The ultimate goal is to check whether \((v, i)\) is contained in \(X(G,C)\).
  In the process, we check for all \((v', i')\) from the DAG \(H_{v, i}\)
  whether they are contained in \(X\).
  We want that \(G \models \psi^{h,n}_{t0, i'}(v')\)
  if and only if we can verify \((v', i') \in X(G,C)\)
  with recursion depth \(h\).
  This corresponds to a node \(t\) of type \(0\) with height at most \(h\) in the tree \(T_{v, i}\)
  with \(v(t) = (v', i')\) and \(W(t) = \varnothing\),
  and \((v', i')\) is in \(X(G,C)\).
  We want that \(G \models \psi^{h,n}_{t1, i', j, c}(v', w)\)
  if and only if we can verify \((v', i') \in X(G,C)\)
  with recursion depth \(h\)
  while stopping the recursion whenever we reach \((w, j)\).
  In these cases, we assume that \((w, j)\) has exactly \(c\) children
  that are contained in \(X(G,C)\).
  This corresponds to a node \(t\) of type \(1\) with height at most \(h\) in \(T_{v, i}\)
  with \(v(t) = (v', i')\), \(W(t) = \set{(w, j)}\),
  and \((v', i') \in X(G,C)\) holds if \((v', i')\) has exactly \(c\) children in \(H_{v, \ell}\)
  that are contained in \(X(G,C)\).
  
  Since, for every \(i' < 1\), \((v', i')\) is not contained in \(X(G,C)\),
  we can already set
  \(\psi^{h,n}_{t0, i'}(x) \deff \bot\) and \(\psi^{h,n}_{t1, i', j, c}(x,y) \deff \bot\)
  for all \(h \in \N\), \(i' \in \Z_{\leq 0}\), \(j \in \Z\), and \(c \in \N\).
  Moreover, for all \(h \in \N\), \(i' \in \Z\), \(j \in \Z_{\leq 0}\), and \(c \in \N\),
  we set \(\psi^{h,n}_{t1, i', j, c}(x,y) \deff \bot\).

  \paragraph*{Preparation}
  Before proceeding, let us introduce a few formulae.
  For all \(d \in \N\), let \(\deg^-_d(x) \deff \exists^{=d}\, y \bigl(E(y,x)\bigr)\).
  Then, \(G \models \deg^-_d(v)\) if and only if \(\deg^-_G(v) = d\).
  Further, for all \(\ell, \ell' \in \N\), we inductively define \(\textsf{path}^{0,n}_{\ell, \ell'}(x,y) \deff (x=y)\) if \(\ell = \ell'\) and
  \begin{align*}
    \textsf{path}^{0,n}_{\ell, \ell'}(x,y)
    = E(x,y) \land \Lor_{d \in [n], \floor*{\frac{\ell-1}{d}} = \ell'} \deg^-_d(y)
  \end{align*}
  else, and, for $h \ge 1$,
  \[\textsf{path}^{h,n}_{\ell, \ell'}(x,y)
    = \exists\, z \Lor_{j=\ell'}^\ell \Bigl[\textsf{path}^{h-1,n}_{\ell, j}(x,z)
    \land \textsf{path}^{h-1,n}_{j, \ell'}(z,y) \Bigr].\]
  We have \(G \models \textsf{path}^{h,n}_{\ell, \ell'}(v,w)\) if and only if
  there is a path in \(H_{v, i}\)
  (and thus also in any other \(H_{v', j}\) that includes \((v, i)\))
  from \((v, \ell)\) to \((w, \ell')\)
  that can be verified in \(h\) recursion steps.

  \paragraph*{Formulae of type \(0\)}
  First, we construct formulae corresponding to nodes of type \(0\).
  Let \(T\) be a tree according to the proof of \cref{lem:5}
  and let $t \in V(T)$ be a node of type \(0\) of the form
  $v(t) = (v',i')$ and $W(t) = \varnothing$.

  In the base case \(h = 0\), where we do not have any further recursion steps left,
  we check that \(t\) is a leaf in \(T\).
  This is the case if \(v'\) does not have any successors \(w\) in \(G\)
  with \(\floor{\frac{i'-1}{\deg^-(w)}} \geq 1\),
  which is equivalent to \(i'-1 \geq \deg^-(w)\).
  For a leaf \(t\) as described above, we have \((v',i') \in X(G,C)\) if and only if \(0 \in C(v')\).
  Thus, for all \(i' \in \Npos\), we set
  \[\psi^{0,n}_{t0, i'}(x) \deff P_0(x) \land \forall\,y \left[\ E(x,y)
    \rightarrow \Lor_{d=i}^n \deg^-_d(y)\ \right].\]

  In the recursion step for \(h \in \Npos\),
  there is some vertex \(a \in V(G)\) and a number \(\ell < i'\)
  such that \((v', i') \dagsle_{H_{v, i}} (a, \ell)\),
  \ie, \((a, \ell)\) is below \((v', i')\) in \(H_{v, i}\).
  Intuitively, the node \((a, \ell)\) should split the DAG \(H_{v, i}\) into parts of almost equal size.
  We guess the number \(c\) of children of \((a, \ell)\) in \(H_{v, i}\) that are in \(X(G,C)\).
  Then, we verify that exactly this number of children is contained in \(X(G,C)\)
  via formulae of type \(0\) and \(h-1\) remaining recursion steps.
  Using the number \(c\), we can verify \((v', i') \in X(G,C)\) with a formula of type \(1\)
  and \(h-1\) remaining recursion steps by passing the information that exactly
  \(c\) children of \((a, \ell)\) are contained in \(X(G,C)\).

  Hence, for \(h, i' \in \Npos\), we set
  \begin{align*}
    \psi^{h,n}_{t0, i'}(x)
    \deff &\psi^{0,n}_{t0, i'}(x)
     \lor \exists\,y \Lor_{\ell=1}^{i'-1} \Lor_{c=0}^n \biggl[\ \textsf{path}^{h,n}_{i',\ell}(x,y)\\
    & \land\ \textsf{children}^{h-1,n}_{t0,\ell,c}(y)
    \land \psi^{h-1,n}_{t1, i', \ell, c}(x, y)\ \biggr]
    \intertext{with}
    \textsf{children}^{h,n}_{t0,\ell,c}(y) \deff &\\ 
    &\hspace*{-6em}\exists^{=c}\,z \left[\ E(y,z) \land \Lor_{d=1}^n \left(\deg^-_d(z)
    \land \psi^{h,n}_{t0, \floor{\frac{\ell-1}{d}}}(z)\right)\ \right]
  \end{align*}
  for all \(h \in \N, \ell \in \Npos\), and \(c \in \N\),
  expressing that ``$(y,\ell)$ admits $c$ children of type $0$ which lie in $X(G,C)$.
  This can be verified in $h$ recursion steps.''

  \paragraph*{Formulae of type \(1\)}
  Now, we construct formulae corresponding to nodes of type \(1\).
  Let \(T\) be a tree according to the proof of \cref{lem:5}
  and let $t \in V(T)$ be a node of type \(1\) of the form
  $v(t) = (v',i')$ and $W(t) = \set{(w,j)}$.

  In the case \(h = 0\), where we do not have any further recursion steps left,
  we check that \(t\) is a leaf in \(T\).
  This happens if \(v' = w\) and \(i' = j\).
  For such a leaf, assuming that exactly \(c\) children of \((v', i')\) are in \(X(G,C)\),
  we have \((v', i') \in X(G,C)\) if and only if \(c \in C(v')\).
  Thus, for all \(i' \in \Npos\) and \(c \in \N\), we set
  \(\psi^{0,n}_{t1, i', i', c}(x,y) \deff P_c(x) \land x=y\).
  Furthermore, for all \(i', \ell \in \Npos\) and \(c \in \N\)
  with \(i' \neq \ell\), we set \(\psi^{0,n}_{t1, i', \ell, c}(x,y) \deff \bot\).

  In the recursion step for \(h \in \Npos\),
  there is some vertex \(a \in V(G)\) and an \(\ell\) with \(j < \ell < i'\)
  such that \((v', i') \dagsle_{H_{v, i}} (a, \ell) \dagsle_{H_{v, i}} (w, j)\).
  We guess the number \(c'\) of children of \((a, \ell)\) in \(H_{v, i}\) that are in \(X(G,C)\).
  Then, we verify that exactly this number is contained in \(X(G,C)\).
  If the child is not above \((w, j)\) in \(H_{v, i}\), then it is a node of type \(0\),
  and we use a formula of type \(0\) with \(h-1\) remaining recursion steps.
  If the child is above \((w, j)\) in \(H_{v, i}\), then it is a node of type \(1\),
  and we use a formula of type \(1\) with \(h-1\) remaining recursion steps,
  passing the information that exactly \(c\) children of \((w,j)\) are contained in \(X(G,C)\).
  Then, using the guessed number \(c'\),
  we can verify \((v', i') \in X(G,C)\) with a formula of type \(1\)
  and \(h-1\) remaining recursion steps by passing the information that exactly
  \(c'\) children of \((a, \ell)\) are contained in \(X(G,C)\).  

  Hence, for all \(h, i', j \in \Npos\) and \(c \in \N\), we set
  \begin{align*}
    \psi^{h,n}_{t1, i', j, c}(x,y) \deff & \psi^{0,n}_{t1, i', j, c}(x,y) \\ 
    &\hspace*{-6em}\lor \exists\,z \Lor_{\ell=j+1}^{i'-1} \Lor_{c'=0}^n \biggl(
      \textsf{path}^{h,n}_{i',\ell}(x,z) \land \textsf{path}^{h,n}_{\ell,j}(z,y)\\
    &\hspace*{-4em}\land \textsf{children}^{h-1,n}_{t1, \ell, j ,c ,c'}(z,y)
      \land \psi^{h-1,n}_{t1, i', \ell, c'}(x, z)\biggr)
    \intertext{with}
    \textsf{children}^{h,n}_{t1, \ell, j ,c, c'}(z,y) \deff &\\
    &\hspace{-8em}\exists^{=c'}\,z' \Biggl(E(z,z') \land  \Lor_{d=1}^n \biggl[\deg^-_d(z')\\
    &\hspace{-8em}\land \quad \biggr(\left[\psi^{h,n}_{t0, \floor{\frac{\ell-1}{d}}}(z')
      \land \neg \textsf{path}^{h,n}_{\floor{\frac{\ell-1}{d}},j}(z',y)\right] \\
    &\hspace{-8em}\lor \quad \left[\psi^{h,n}_{t1, \floor{\frac{\ell-1}{d}}, j, c}(z',y)
      \land \textsf{path}^{h,n}_{\floor{\frac{\ell-1}{d}},j}(z',y)\right]
      \biggr)\biggr]\Biggr)
  \end{align*}
  for all \(h \in \N\), \(\ell, j \in \Npos\), and \(c, c' \in \N\),
  expressing that ``$(z,\ell)$ admits $c'$ children which lie in $X(G,C)$
  if \(c\) children of \((y, j)\) are contained in \(X(G,C)\).
  The children of \((z,\ell)\) above \((y, j)\) are of type \(1\)
  and those not above \((y, j)\) are of type \(0\).
  All of this can be verified in $h$ recursion steps''

  \paragraph*{Nesting depth of the formulae}
  Since, by \cref{lem:5}, there is a tree \(T_{v, i}\) that,
  by \cref{clm:tree-height}, has height \((4r+2) \cdot \log (n+1)\),
  it suffices to have formulae \(\psi^{h,n}_{t0, i}(x)\) with logarithmic nesting depth.
  That is, we choose
  \[\phi^{(n)}_i(x) \deff \psi^{((4r+2) \cdot \log (n+1)),n}_{t0, i}(x).\]

  Then, \(\A(G,C) \models \phi^{(n)}_i (v)\) if and only if \((v, i) \in X(G,C)\)
  for all graphs \(G\) of size \(\abs{G} \leq n\),
  all cardinality conditions \(C\) for \(G\),
  and all \(v \in V(G)\).
\end{IEEEproof}

We are ready to prove the main result of this section.

\lreqeqtock*

\begin{IEEEproof}
  We proceed by induction on the structure of \(\phi\).
  For formulae \(\phi(\bar{x}, \bar{\kappa}) \in \FOC[\tau]\),
  since we only need equivalence on structures of size at most \(n\),
  we can apply the arguments from the proof of \cite[Proposition 8.4.18]{ebbinghaus2005finite},
  first replacing $\#$-operators by counting quantifiers, then hard-coding families of formulae inductively,
  beginning with $\top$ resp.\ $\bot$ for atomic number sentences ($\le, \min, \max, S$) and then replacing existential numeric quantification by disjunctions over $N(\A)$ for every possible assignment of the previously quantified variable.
  Hence, there are constants \(k, r \in \N\) and,
  for every \(n \in \N\),
  a family of \(\Ck{r}[\tau]\)-formulae
  \(\bigl(\psi_{\bar{j}}(\bar{x})\bigr)_{\bar{j} \in [n]^{\abs{\bar{\kappa}}}}\)
  such that for all \(\tau\)-structures \(\A\) of size \(\abs{\A} \leq n\),
  all \(\bar{v} \in \bigl(V(\A)\bigr)^{\abs{\bar{x}}}\),
  and all \(\bar{j} \in [\abs{\A}]^{\abs{\bar{\kappa}}}\), it holds that
  \[\A^+ \models \phi(\bar{v}, \bar{j}) \iff \A \models \psi_{\bar{j}}(\bar{v}).\]

  For \(\phi = (\neg \phi_1)\),
  \(\phi = (\phi_1 \lor \phi_2)\),
  \(\phi = (\exists y \, \phi_1)\),
  \(\phi = (\# \iota \, \phi_1 = \kappa)\), or
  \(\phi = (\# y \, \phi_1 = \kappa)\),
  where
  \begin{itemize}
      \item \(y\) is a domain variable, 
      \item \(\iota, \kappa\) are number variables,
      \item \(\bar{x}\) is a tuple of domain variables,
      \item \(\bar{\kappa}\) is a tuple of number variables we assume to contain \(\kappa\),
      \item \(\free(\phi) \subseteq \bar{x} \cup \bar{\kappa}\), and
      \item \(\phi_1, \phi_2\) are \(\LRECeq\)-formulae,
  \end{itemize}
  we first construct families of \(\logCk\)-formulae
  \(\bigl(\psi_{1,\bar{j}}(\bar{x})\bigr)_{\bar{j} \in [n]^{\abs{\bar{\kappa}}}}\),
  \(\bigl(\psi_{2,\bar{j}}(\bar{x})\bigr)_{\bar{j} \in [n]^{\abs{\bar{\kappa}}}}\)
  recursively and then again apply the arguments from the proof of \cite[Proposition 8.4.18]{ebbinghaus2005finite}.

  Now let
  \(\phi = \bigl[\lrec_{\bar{y}_1, \bar{y}_2, \bar{\iota}}\, \phi_=, \phi_\texttt{E}, \phi_\texttt{C}\bigr](\bar{x}, \bar{\kappa})\)
  for some compatible tuples of domain variables \(\bar{y}_1, \bar{y}_2, \bar{x}\),
  non-empty tuples of number variables \(\bar{\iota}, \bar{\kappa}\),
  and \(\LRECeq\)-formulae
  \(\phi_=(\bar{y}_1, \bar{y}_2, \bar{x})\),
  \(\phi_\texttt{E}(\bar{y}_1, \bar{y}_2, \bar{x})\),
  \(\phi_\texttt{C}(\bar{y}_1, \bar{x}, \bar{\iota}, \bar{\kappa})\).

  By the induction hypothesis, there is a constant \(k' \in \N\)
  such that for every \(n \in \N\),
  there are \(\C{k'}{\bigO{\log n}}\)-formulae
  \(\psi_=(\bar{y}_1, \bar{y}_2, \bar{x}), \psi_\texttt{E}(\bar{y}_1, \bar{y}_2, \bar{x})\)
  as well as a family of \(\C{k'}{\bigO{\log n}}\)-formulae
  \(\bigl(\psi_{\texttt{C},\bar{i}\bar{j}}(\bar{y}_1, \bar{x})\bigr)_{\bar{i}\bar{j}
  \in [n]^{\abs{\bar{\iota}\bar{\kappa}}}}\)
  with
\begin{align*}
    \A^+ \models \phi_=(\bar{u}_1, \bar{u}_2, \bar{v})
  &\iff \A \models \psi_=(\bar{u}_1, \bar{u}_2, \bar{v}), \\
  \A^+ \models \phi_\texttt{E}(\bar{u}_1, \bar{u}_2, \bar{v})
  &\iff \A \models \psi_\texttt{E}(\bar{u}_1, \bar{u}_2, \bar{v})\text{, and} \\
  \A^+ \models \phi_\texttt{C}(\bar{u}_1, \bar{v},\bar{i}, \bar{j})
  &\iff \A \models \psi_{\texttt{C},\bar{i}\bar{j}}(\bar{u}_1, \bar{v})
\end{align*}
  for all structures \(\A\) of size \(\abs{\A} \leq n\)
  and all \(\bar{u}_1 \in \bigl(V(\A)\bigr)^{\abs{\bar{y}_1}}\),
  \(\bar{u}_2 \in \bigl(V(\A)\bigr)^{\abs{\bar{y}_2}}\),
  \(\bar{v} \in \bigl(V(\A)\bigr)^{\abs{\bar{x}}}\),
  \(\bar{i} \in N(\A)^{\abs{\bar{\iota}}}\), and
  \(\bar{j} \in N(\A)^{\abs{\bar{\kappa}}}\).
  Moreover, by \cref{theo:Ck-for-X},
  there is a \(k'' \in \Npos\) such that for all \(n \in \Npos\) and
  \(\ell \in \bigset{1, \dots, (n+1)^{\abs{\bar{\kappa}}}}\),
  there is a \(\C{k''}{\bigO{\log n}}[\set{E, P_0, \dots, P_n}]\)-formula
  \(\phi_{X,\ell}(x)\) such that for all \(\tau\)-structures \(\A\)
  of size at most \(n\),
  and for \(G = (V,E)\) and \(C\) from the \(\LRECeq\) definition
  in \cref{sec:preliminaries} for \(\phi\),
  it holds that
  \(\A(G,C) \models \phi_{X,\ell}(v) \iff (v,\ell)\in X(G,C)\) for all \(v \in V(G)\).
  We turn these into \(\logCk[\tau]\)-formulae \(\psi_{X,\ell}(\bar{x})\) by
  replacing every occurrence of
  \begin{itemize}
    \item \(z_1 = z_2\) by
    \vspace{-1ex}
      \[\exists\, z_{1,1} \cdots \exists\, z_{1,\abs{\bar{x}}}
        \exists\, z_{2,1} \cdots \exists\, z_{2,\abs{\bar{x}}}
        \bigl(\psi_=(\bar{z_1}, \bar{z_2}, \bar{x})\bigr)\]
      with \(\bar{z}_i = (z_{i,1}, \dots, z_{i,\abs{\bar{x}}})\) for \(i \in \set{1,2}\),
    \item \(\exists z_1\) by \(\exists z_{1,1} \dots \exists z_{1,\abs{x}}\),
    \item \(E(z_1,z_2)\) by
      \begin{align*}
        &\exists\, z_{1,1}' \cdots \exists\, z_{1,\abs{\bar{x}}}'
          \exists\, z_{2,1}' \cdots \exists\, z_{2,\abs{\bar{x}}}' \bigl(\\
        &\psi_=(\bar{z}_1', \bar{z}_1, \bar{x})
          \land \psi_=(\bar{z}_2', \bar{z}_2, \bar{x})
          \land \psi_\texttt{E}(\bar{z}_1', \bar{z}_2', \bar{x})\bigr),
      \end{align*}
    \item and \(P_i(z_1)\) by \(\exists\, z_{1,1}' \cdots \exists\, z_{1,\abs{\bar{x}}}'
      \bigl(\psi_=(\bar{z}', \bar{z}, \bar{x}) \land \psi_{\texttt{C},\bar{i}\bar{\ell}}(\bar{z}', \bar{x})\bigr)\)
  \end{itemize}
  for any variables \(z_1, z_2\) that occur in \(\phi_{X, \ell}\).

  Let \(\psi_{\bar{m}}(\bar{x}) \deff \psi_{X, \angles{\bar{m}}}(\bar{x})\)
  for all \(\bar{m} \in [n]^{\abs{\bar{\kappa}}}\).
  Then, for all \(\tau\)-structures \(\A\) of size \(\abs{\A} \leq n\),
  all \(\bar{v} \in \bigl(V(\A)\bigr)^{\abs{\bar{x}}}\),
  and all \(\bar{m} \in [\abs{\A}]^{\abs{\bar{\kappa}}}\), it holds that
  \begin{align*}
    \A^+ \models \phi(\bar{v}, \bar{m})
    &\iff (\bar{v}, \angles{\bar{m}}) \in X(G,C)\\
    &\iff \A \models \psi_{X, \angles{\bar{m}}}(\bar{v})\\
    &\iff \A \models \psi_{\bar{m}}(\bar{v}).
  \end{align*}
\end{IEEEproof}
 
\section{Interval Graphs}
\label{sec:interval-graphs}
\begin{figure}
    \centering
    \begin{subfigure}[c]{.49\textwidth}
        \centering
        \begin{tikzpicture}
            \node[] (ax) at (-.5,6) {$\dots$};
            \node[] (a0) at (-.5,6) {};
            \node[] (a1) at (2,6) {};
        
            \node[] (b0) at (0,5.5) {};
            \node[] (b1) at (2.5,5.5) {};
            \node[] (bx) at (2.5,5.5) {$\dots$};

            \node[] (c0) at (0,4.5) {};
            \node[] (c1) at (.5,4.5) {};

            \node[] (d0) at (0,5) {};
            \node[] (d1) at (1.5,5) {};

            \node[] (e0) at (0.5,4.5) {};
            \node[] (e1) at (2,4.5) {};

            \node[] (f0) at (1.5,5) {};
            \node[] (f1) at (2,5) {};

            \node[] (g0) at (0.5,4) {};
            \node[] (g1) at (1,4) {};

            \node[] (h0) at (1,4) {};
            \node[] (h1) at (1.5,4) {};

            \draw[-,very thick,draw=black] (a0)  -- node [midway,below] {$a$} (a1);
            \draw[-,very thick,draw=black] (b0) -- node [midway,below] {$b$} (b1);
            \draw[-,very thick,draw=marine] (c0) -- node [midway,below] {$\textcolor{marine}{c}$} (c1);
            \draw[-,very thick,draw=marine] (d0) -- node [midway,below] {$\textcolor{marine}{d}$} (d1);
            \draw[-,very thick,draw=marine] (e0) -- node [midway,below] {$\textcolor{marine}{e}$} (e1);
            \draw[-,very thick,draw=marine] (f0) -- node [midway,below] {$\textcolor{marine}{f}$} (f1);
            \draw[-,very thick,draw=red] (g0) -- node [midway,below] {$\textcolor{red}{g}$} (g1);
            \draw[-,very thick,draw=red] (h0) -- node [midway,below] {$\textcolor{red}{h}$} (h1);
        \end{tikzpicture}
    \end{subfigure}
    \begin{subfigure}[c]{.49\textwidth}
        \centering
        \begin{tikzpicture}
            \node[] (1) at (-1,10) {$\ddots$};
            \node[fill=black,ultra thick,circle,inner sep=2pt,label={above:$a$}] (2) at (0,9) {};
            \node[fill=black,ultra thick,circle,inner sep=2pt,label={above:$b$}] (3) at (1.5,9) {};
            \node[] (4) at (2.5,10) {\reflectbox{$\ddots$}};

            \node[fill=marine,ultra thick,circle,inner sep=2pt,label={below:$\textcolor{marine}{c}$}] (5) at (-1.5,7.5) {};
            \node[fill=marine,ultra thick,circle,inner sep=2pt,label={below:$\textcolor{marine}{d}$}] (6) at (-0.5,7) {};
            \node[fill=marine,ultra thick,circle,inner sep=2pt,label={below:$\textcolor{marine}{e}$}] (7) at (2,7) {};
            \node[fill=marine,ultra thick,circle,inner sep=2pt,label={below:$\textcolor{marine}{f}$}] (8) at (3,7.5) {};

            \node[fill=red,ultra thick,circle,inner sep=2pt,label={below:$\textcolor{red}{g}$}] (9) at (0,5.5) {};
            \node[fill=red,ultra thick,circle,inner sep=2pt,label={below:$\textcolor{red}{h}$}] (10) at (1.5,5.5) {};

            \node[] (d) at (0,4.5) {};
            
            \draw[-] (1) -- (2);
            \draw[-] (2) -- (3);
            \draw[-] (3) -- (4);

            \foreach \x in {2,3}
                \foreach \y in {5,...,8}{
                    \draw[-,blue] (\x) -- (\y);}
            \foreach \x in {2,3,6,7}
                \foreach \y in {9,10}{
                    \draw[-,orange] (\x) -- (\y);}
                    
            \draw[-,marine] (5) -- (6);
            \draw[-,marine] (6) -- (7);
            \draw[-,marine] (7) -- (8);
        \end{tikzpicture}
    \end{subfigure}
    \caption{Part of an interval representation (left) and its interval graph $G$ admitting nested modules (right). 
    The four (visible) maxcliques $C_{i,i \in [4]}$ of $G$ are 
    $\set{\textcolor{black}{a},\textcolor{black}{b},\textcolor{marine}{c},\textcolor{marine}{d}}$,
    $\set{\textcolor{black}{a},\textcolor{black}{b},\textcolor{marine}{d},\textcolor{marine}{e},\textcolor{red}{g}}$,
    $\set{\textcolor{black}{a},\textcolor{black}{b},\textcolor{marine}{d},\textcolor{marine}{e},\textcolor{red}{h}}$, and
    $\set{\textcolor{black}{a},\textcolor{black}{b},\textcolor{marine}{e},\textcolor{marine}{f}}$.}
    \label{fig:interval_representation}
\end{figure}
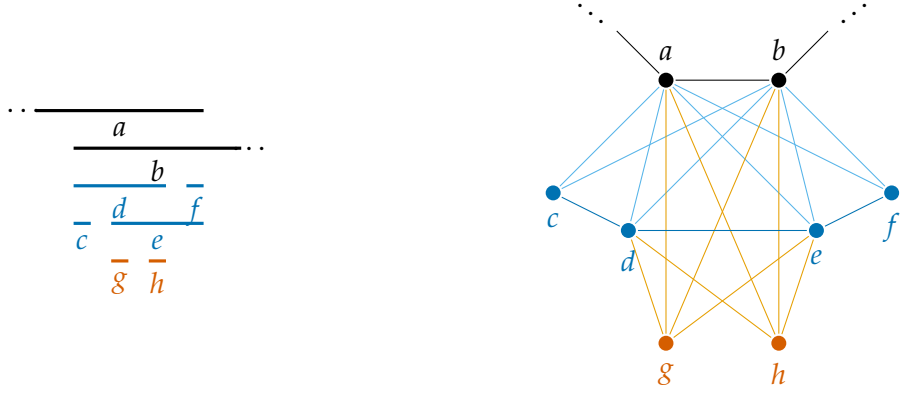 
In this section, we describe how the result from \cref{theo:lreceq_to_c_k} allows us to obtain, from an \LRECeq-definable canonisation of interval graphs, a $k \in \N$ such that, for every $n \in \N$, $\logCk$ identifies every interval graph of order $n$. We obtain a similar result for chordal claw-free graphs.
Finally, we sketch how the result we obtained for interval graphs can be shown without the need for \LRECeq, using an \STCC-definable canonisation for a subclass of interval graphs and the fact that every interval graph can be decomposed into interval graphs of that subclass.

An \emph{interval} is a set of consecutive integers. An \emph{interval representation} $\mathcal{I}$ is a set of intervals, from which we get its graph $G_{\mathcal{I}}$ with $V(G_{\mathcal{I}}) \coloneqq \mathcal{I}$ and $E(G_{\mathcal{I}}) \coloneqq \set{\set{I,J} \subseteq \mathcal{I} \mid I \cap J \neq \varnothing}$.
An undirected graph $G$ is an \emph{interval graph} if there exists an interval representation $\mathcal{I}$ such that $G \cong G_{\mathcal{I}}$,
see \cref{fig:interval_representation} for an example.
An interval representation $\mathcal{I}$ is \emph{(cardinalitywise) minimal} if $\bigcup \mathcal{I} \subset \N$ is minimal with respect to all interval representations $\mathcal{I}'$ with $G_{\mathcal{I}} \cong G_{\mathcal{I'}}$.
An interval graph $G$ is \emph{proper} if there is an interval representation $\mathcal{I}$ with $G_{\mathcal{I}} \cong G$ and, for all $I,J \in \mathcal{I}$, $I \not \subseteq J$.

\begin{lemma}[\cite{laubner2013recursion}]\label{lemm:inteval_lrec}
  There exists an $\LRECeq$-definable canonisation of interval graphs $\psi(\iota,\kappa)$ such that, for all $n \in \N$ and interval graphs $G$ of order $n$, it holds that
  $\mu\colon G^+ \cong ([n],\psi[G^+;\iota,\kappa])$.
\end{lemma}

In particular, since $\psi$ is the result of a canonisation, it holds for all $n \in \N$ and interval graphs $G$ and $H$ of order $n$ that $G \cong H$ iff \(([n],\psi[G^+;\iota,\kappa]) = ([n],\psi[H^+;\iota,\kappa]).\)

Applying \cref{theo:lreceq_to_c_k}, we obtain:

\begin{corollary}
  There exists a $k \in \N$ such that for all $n \in \N$ 
  there is a family of $\C{k}{\bigO{\log n}}$-sentences $\left(\psi_{ij}\right)_{i,j \in [n]}$ such that for all interval graphs $G$ of order $n$ and $i,j \in [n]$
  \[G \models \psi_{ij} \iff \; G^+ \models \psi(i,j).\]
\end{corollary}

Now, let, for every $n \in \N$, interval graph $G$ of order $n$ and all $i,j \in [n]$,
\[
\psi^G_{ij} \coloneqq \begin{cases}
                          \top & \text{if } G \models \psi_{ij}, \\
                          \bot & \text{if } G \not\models \psi_{ij}.
                      \end{cases}
\]
\begin{lemma}\label{lemm:interval_interval_separation}
    There exists a $k \in \N$ such that, for every $n \in \N$, every interval graph $G$ of order $n$ admits a $\C{k}{\bigO{\log n}}$-formula $\phi_G$ satisfying, for every interval graph $H$ of order $n$,
    \[H \models \phi_G \iff H \cong G.\]
\end{lemma}

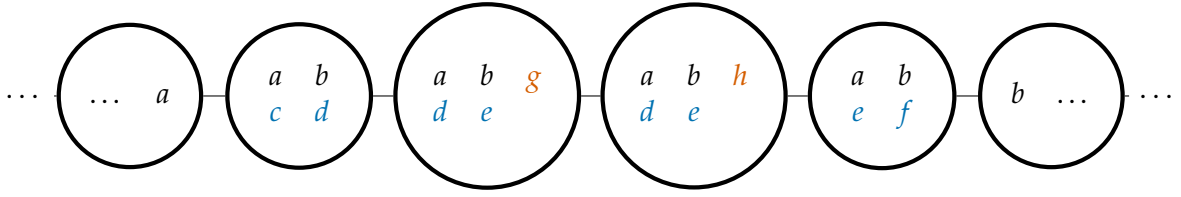
\begin{figure*}[t!]
    \centering
        \begin{tikzpicture}
            \node[] (0) at (-1.4,0) {$\dots$};
            \node[circle, draw=black, ultra thick] (1) at (0,0) {
            \begin{tabular}{cc}
                \dots & $\textcolor{black}{a}$ \\
            \end{tabular}
            };
            \node[circle, draw=black, ultra thick] (2) at (2.25,0) {
            \begin{tabular}{cc}
                $\textcolor{black}{a}$ & $\textcolor{black}{b}$ \\
                $\textcolor{marine}{c}$ & $\textcolor{marine}{d}$
            \end{tabular}
            };
            \node[circle, draw=black, ultra thick] (3) at (4.75,0) {
            \begin{tabular}{ccc}
                $\textcolor{black}{a}$ & $\textcolor{black}{b}$ & $\textcolor{red}{g}$ \\
                $\textcolor{marine}{d}$ & $\textcolor{marine}{e}$ &
            \end{tabular}
            };
            \node[circle, draw=black, ultra thick] (4) at (7.5,0) {
            \begin{tabular}{ccc}
                $\textcolor{black}{a}$ & $\textcolor{black}{b}$ & $\textcolor{red}{h}$ \\
                $\textcolor{marine}{d}$ & $\textcolor{marine}{e}$ &
            \end{tabular}
            };
            \node[circle, draw=black, ultra thick] (5) at (10,0) {
            \begin{tabular}{cc}
                $\textcolor{black}{a}$ & $\textcolor{black}{b}$ \\
                $\textcolor{marine}{e}$ & $\textcolor{marine}{f}$
            \end{tabular}
            };
            \node[circle, draw=black, ultra thick] (6) at (12.25,0) {
            \begin{tabular}{cc}
                $\textcolor{black}{b}$ & \dots \\
            \end{tabular}
            };
            \node[] (7) at (13.675,0) {$\dots$};

            \draw[] (0) -- (1);
            \draw[] (1) -- (2);
            \draw[] (2) -- (3);
            \draw[] (3) -- (4);
            \draw[] (4) -- (5);
            \draw[] (5) -- (6);
            \draw[] (6) -- (7);
        \end{tikzpicture}
    \caption{Part of a possible path decomposition for the graph $G$ from \cref{fig:interval_representation}. In particular, observe that other possible path decompositions can be obtained by totally reversing the contents of the four, resp.\ two centremost nodes.}
    \label{fig:interval_path}
\end{figure*}
 
\begin{IEEEproof}
    Let $G$ be an interval graph of order $n$. We claim that
    \[\phi^G \coloneqq \bigwedge_{i,j \in [n]} \psi_{ij} \leftrightarrow \psi^G_{ij}\]
    is that formula.
    To that end, let $H$ be an interval graph of order $n$.

    $(\!{\implies}\!)$ Suppose that $H \models \phi_G$.
    Then, for all $i,j \in [n]$,
    $H \models \psi_{ij} \leftrightarrow \psi^G_{ij}$,
    implying that $H \models \psi_{ij}$ iff $G \models \psi_{ij}$.
    Thus, $([n],\psi[G^+;\iota,\kappa]) = ([n],\psi[H^+;\iota,\kappa])$ and hence, $G \cong H$.

    $(\!{\impliedby}\!)$ Suppose that $G \cong H$.
    Then, $([n],\psi[G^+;\iota,\kappa]) = ([n],\psi[H^+;\iota,\kappa])$.
    Thus, for all $i,j \in [n]$, $H \models \psi_{ij}$ iff $G \models \psi_{ij}$
    and thus $\psi_{ij} \leftrightarrow \psi^G_{ij}$.
    Hence, $H \models \phi_G$.
\end{IEEEproof}

It thus remains to show that we can separate an interval graph from those that are not interval graphs or are of a different order.
For this, we need the logics $\STC$ and $\STCC$.

\emph{Symmetric transitive closure logic} \STC (see \cite{laubner2011structure}) extends \FO by the \stc\emph{-operator},
which, for all vocabularies $\tau$, $\tau$-structures $\A$ and $k \in \N$,
allows the definition of an undirected graph over vertex $k$-tuples.
Syntactically, if $\psi$ is an $\STC[\tau]$-formula,
$\bar{x}$ and $\bar{y}$ are $k$-tuples of variables, and $\bar{v}, \bar{w} \in V(\A)^k$, then
\(\phi \coloneqq [\stc_{\bar{x},\bar{y}} \psi](\bar{v},\bar{w})\)
is also an $\STC$-formula.
Concerning the semantics, it suffices for us to know that in $\phi$,
$\psi$ defines an undirected graph over $V(\A)^k$.
The \stc-operator then tests whether $(\bar{v},\bar{w})$ is an edge in the symmetric transitive closure of said graph.
In that sense, it is very close to, though more restrictive than,
the \textsf{tc}-operator from transitive closure logic \cite{immerman1999descriptive}.
The extension of $\STC$ to two-sorted structures then yields $\STCC$,
where the $\stc$-operator is extended over mixed domain/number tuples.
$\STCC$ has been found to be contained in $\LRECeq$.

\begin{lemma}[\cite{laubner2013recursion}]\label{lemm:LRECeqtoSTCC}
    $\STCC \le \LRECeq$.
\end{lemma}

Thus, applying \cref{theo:lreceq_to_c_k}, we immediately obtain the following.

\begin{lemma} \label{lemm:stcc_to_c_k}
  For every vocabulary \(\tau\)
  and every \(\STCC[\tau]\)-formula \(\phi(\bar{x}, \bar{\iota})\),
  there is a constant \(k \in \N\) such that for every \(n \in \N\),
  there is a family of \(\logCk\)-formulae
  \(\bigl(\psi_{\bar{i}}(\bar{x})\bigr)_{\bar{i} \in [n]^{\abs{\bar{\iota}}}}\)
  such that for all \(\tau\)-structures \(\A\) of size \(\abs{\A} \leq n\),
  all \(\bar{v} \in \bigl(V(\A)\bigr)^{\abs{\bar{x}}}\),
  and all \(\bar{i} \in (N(\A))^{\abs{\bar{\iota}}}\), it holds that
  \[\A^+ \models \phi(\bar{v}, \bar{i}) \iff \A \models \psi_{\bar{i}}(\bar{v}).\]
\end{lemma}

It is worth mentioning that we can also obtain the above result without passing through \LRECeq.
The approach is similar to the one of \cref{theo:lreceq_to_c_k},
an induction over the structure of the formula, although easier,
as the \stc-operator can be modelled as a simple connectivity test on a graph defined over vertex/number $k$-tuples.
This can be expressed through a formula of depth logarithmic in the size of the graph (see \cite[Example 3]{grohe2021logarithmic}).

\begin{lemma}[\cite{laubner2011structure}]\label{lemm:interval_stc}
  The class of interval graphs is $\STC$-definable.
\end{lemma}

\begin{corollary}\label{coro:interval_c_k}
  There is a $k \in \N$ such that, for every $n \in \N$,
  there exists a \logCk-sentence $\varphi^{(n)}_{\textsf{interval}}$ such that for every graph $G$ of order $n$, it holds that
  $G \models \varphi^{(n)}_{\textsf{interval}}$
  if and only if $G$ is an interval graph.
\end{corollary}

By combining
\cref{lemm:interval_interval_separation,coro:interval_c_k}, 
we can now prove our main theorem of this section.

\interval*

\begin{IEEEproof}
  Let $H$ be a graph with $G \not \cong H$.
  If $|V(H)| \neq |V(G)|$, then $G$ and $H$ are separated by the formula $\exists^{=n}x \, (x=x)$, where $n$ is the order of $G$.
  Thus, let $|V(H)| = |V(G)|$.
  If $V(H)$ is not an interval graph, then, by \cref{coro:interval_c_k},
  $G$ and $H$ are separated by some formula
  $\varphi^{(n)}_{\textsf{interval}} \in \C{k'}{\bigO{\log n}}[\set{E}]$
  for some fixed $k' \in \N$.
  Therefore, suppose that $H$ is an interval graph of order $n$.
  Then, by \cref{lemm:interval_interval_separation} $H \not \models \phi^G$
  with $\phi^G \in \C{k''}{\bigO{\log n}}[\set{E}]$ for some fixed $k'' \in \N$.
  Hence, letting $k = \max(k',k'')$,
  the formula $\Phi^G \in \C{k}{\bigO{\log n}}[{E}]$ defined as
  \[\Phi^G \coloneqq \exists^{=n}x \, (x=x) \land \varphi^{(n)}_{\textsf{interval}} \land \phi^G\]
  describes $G$ up to isomorphism.
\end{IEEEproof}

\begin{corollary}
  There is a $k \in \N$ such that $\WL{k}{\bigO{\log n}}$ identifies every $n$-vertex interval graph.
\end{corollary}

We obtain a similar result for chordal claw-free graphs.
A graph is \emph{chordal} if every cycle of length at least $4$ admits a chord.
This can be expressed by an $\STC$-sentence which, for every path of length $3$ of a graph $G$,
tests that it cannot be closed to an induced cycle of length at least $4$.
A graph is \emph{claw-free} if it has no induced subgraph isomorphic to the complete bipartite graph $K_{1,3}$.
It is clear that this can be tested by a $\C{4}{4}$-formula.
Thus, whether a graph is chordal and claw-free can be tested by an \STC-sentence.
In addition, chordal claw-free graphs admit \(\LRECeq\)-definable canonisation.

\begin{lemma}[\cite{grussien2019capturing}]
  There exists an $\LRECeq$-definable canonisation of chordal claw-free graphs $\psi(\iota,\kappa)$ such that, for all interval graphs $G$,
  \[\mu \colon G^+ \cong ([n],\psi[G^+;\iota,\kappa]).\]
\end{lemma}

An analogous argumentation yields \cref{theo:claw_free_chordal}.

\clawfreechordal*

\subsection{Circumventing $\LRECeq$}

\begin{figure}[t!]
        \centering
        \begin{tikzpicture}
            \node[rectangle,draw=black,very thick] (1) at (0,10) {
                \begin{tikzpicture}
                    \node[] (1) at (-1,10) {$\ddots$};
                    \node[fill=black,ultra thick,circle,inner sep=2pt,label={above:$\textcolor{black}{a}$}] (2) at (0,9) {};
                    \node[fill=black,ultra thick,circle,inner sep=2pt,label={above:$\textcolor{black}{b}$}] (3) at (1.5,9) {};
                    \node[] (4) at (2.5,10) {\reflectbox{$\ddots$}};
                    \node[fill=marine,ultra thick,circle,inner sep=2pt] (5) at (0.75,8) {};
                    
                    \draw[-] (1) -- (2);
                    \draw[-] (2) -- (3);
                    \draw[-] (3) -- (4);
                    \foreach \x in {2,3}{
                            \draw[-,blue] (\x) -- (5);}
                \end{tikzpicture}
            };
            \node[rectangle,draw=marine,very thick] (2) at (0,7) {
                \begin{tikzpicture}
                    \node[fill=marine,ultra thick,circle,inner sep=2pt,label={below:$\textcolor{marine}{c}$}] (5) at (-1.5,7.5) {};
                    \node[fill=marine,ultra thick,circle,inner sep=2pt,label={below:$\textcolor{marine}{d}$}] (6) at (-0.5,7) {};
                    \node[fill=marine,ultra thick,circle,inner sep=2pt,label={below:$\textcolor{marine}{e}$}] (7) at (2,7) {};
                    \node[fill=marine,ultra thick,circle,inner sep=2pt,label={below:$\textcolor{marine}{f}$}] (8) at (3,7.5) {};

                    \node[fill=red,ultra thick,circle,inner sep=2pt] (9) at (0.75,6) {};

                    \foreach \x in {6,7}{
                            \draw[-,orange] (\x) -- (9);}
                    
                    \draw[-,marine] (5) -- (6);
                    \draw[-,marine] (6) -- (7);
                    \draw[-,marine] (7) -- (8);
                \end{tikzpicture}
            };
            \node[] (2l) at (-3.5,8) {\dots};
            \node[] (2r) at (3.5,8) {\dots};
            \node[rectangle,draw=red,very thick] (3) at (-1,5) {
                \begin{tikzpicture}
                    \node[fill=red,ultra thick,circle,inner sep=2pt,label={below:$\textcolor{red}{g}$}] (9) at (0.75,6) {};
                \end{tikzpicture}
            };
            \node[rectangle,draw=red,very thick] (4) at (1,5) {
                \begin{tikzpicture}
                    \node[fill=red,ultra thick,circle,inner sep=2pt,label={below:$\textcolor{red}{h}$}] (9) at (0.75,6) {};
                \end{tikzpicture}
            };
            \draw[->,very thick] (1) -- (2);
            \draw[->,very thick,shorten >=0.5cm] (1) -- (2l);
            \draw[->,very thick,shorten >=0.5cm] (1) -- (2r);
            \draw[->,very thick] (2) -- (3);
            \draw[->,very thick] (2) -- (4);
        \end{tikzpicture}
    \caption{A simplified sketch of the modular decomposition corresponding to $G$ from \cref{fig:interval_representation}. 
    Letting $M \neq C_{i,i \in [4]}$ be a possible end of $G$, the vertex set $\textcolor{marine}{S} \coloneqq \bigcup_{i \in [4]} C_i \setminus \bigcup (\MC(G) \setminus \set{C_{i,i\in [4]}}) =  \set{\textcolor{marine}{c},\textcolor{marine}{d},\textcolor{marine}{e},\textcolor{marine}{f},\textcolor{red}{g},\textcolor{red}{h}}$ is a module of $G$, whereas the vertex sets $\set{\textcolor{red}{g}}$ and $\set{\textcolor{red}{h}}$ are modules of $G[\textcolor{marine}{S}]$, with $\set{\textcolor{marine}{c},\textcolor{marine}{d}}$ and $\set{\textcolor{marine}{e},\textcolor{marine}{f}}$ as possible ends of $G[\textcolor{marine}{S}]$.}
    \label{fig:interval_graphs}
\end{figure}
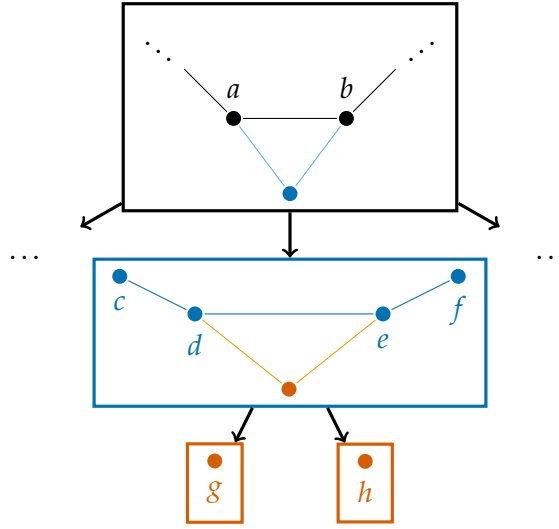 
Finally, we sketch how \cref{theo:interval} can be proved directly
without considering \LRECeq and applying the results due to \citeauthor{laubner2010capturing}~\cite{laubner2010capturing}
and \citeauthor{grussien2019capturing}~\cite{grussien2019capturing}.
We begin with some insights into the properties of interval graphs.
A \emph{maxclique} of a graph $G$ is a vertex subset $C \subseteq V(G)$
such that $C$ forms a clique and, for all $v \in V(G) \setminus C$,
$C \cup \set{v}$ does not form a clique.
By $\MC(G)$, we denote the set of all maxcliques of $G$.

Observe that interval graphs are exactly those graphs
whose maxcliques can be brought into a (not necessarily unique) linear order such that
every vertex is contained in consecutive maxcliques of that order \cite{mohring1985algorithmic}.
Equivalently, they are those graphs admitting path decompositions (see \cite{robertson1983graph})
of which every bag corresponds to a maxclique, as shown in \cref{fig:interval_path}.

A \emph{possible end} of an interval graph $G$ is a maxclique $M \in \MC(G)$
such that $G$ admits a path decomposition $(P,\beta)$ satisfying,
for some $p \in V(P)$ with $\deg_{P}(p) = 1$, that $\beta(p) = M$.
Given such a possible end $M$ of an interval graph $G$,
we can obtain an \STC-definable strict weak order $\prec_{M}$
which captures all the information about the order of the maxcliques \cite{laubner2010capturing}.
Formally, $\prec_M$ is initialised as $M \prec_M C$ for all $C \in \MC(G) \setminus \set{M}$
and is then recursively extended through
\[C \prec_M D \text{ if } \exists X \in \MC(G): \begin{cases}
     & X \prec_M D, (X \cap C) \setminus D \neq \varnothing, \\
     & C \prec_M X, (X \cap D) \setminus C \neq \varnothing. \\
\end{cases}\]

Notably, for some interval graphs $G$ and possible ends $M$,
the order $\prec_M$ becomes a linear order over $\MC(G)$;
it can be shown that extending it over $V(G)$ induces a strict weak order $<_G$
in which two vertices $v, w \in V(G)$ are $<_G$-incomparable iff $N_G[v] = N_G[w]$.
This yields an $\STCC$-definable canonisation for those interval graphs \cite{laubner2010capturing}.
In particular, thanks to \cref{lemm:stcc_to_c_k},
we obtain a fixed $k^*$ and a $\C{k^*}{\bigO{\log n}}$-formula
for each such interval graph describing it up to isomorphism.

Now, we consider those interval graphs $G$ for which $\prec_M$ is not a linear order.
A set $\mathcal{C} \subseteq \MC(G)$ of maxcliques is \emph{incomparable} wrt.\ $\prec_{M}$
if for all pairwise distinct maxcliques $C_1, C_2 \in \mathcal{C}$,
neither $C_1 \prec_M C_2$ or $C_2 \prec_M C_1$.
A set \(\mathcal{C} \subseteq \MC(G)\) is \emph{maximal} if,
for all $C_1 \in \mathcal{C}$ and $C \in \MC(G) \setminus \mathcal{C}$, $C_1 \prec_M C$ or $C \prec_M C_1$.
For each such maximal incomparable set of maxcliques $\mathcal{C} \subseteq \MC(G)$
and $\mathcal{D} \coloneqq \MC(G) \setminus \mathcal{C}$,
the vertex set $S_\mathcal{C}\coloneqq \bigcup \mathcal{C} \setminus \bigcup \mathcal{D}$ is a \emph{module},
a vertex subset with a uniform connectivity behaviour towards the rest of the graph,
that is, for every $v \in V(G) \setminus S_\mathcal{C}$,
either $vw \in E(G)$ or $vw \not\in E(G)$ for all $w \in S_\mathcal{C}$.
In the following, we consider only those modules that are obtained through maximal incomparable sets of maxcliques.
Note that for each module $S_\mathcal{C}$, it holds that $G[S_\mathcal{C}]$ is also an interval graph, see \cref{fig:interval_graphs}.

There, we see how inductively replacing modules by individual vertices
(which maintain the same connectivity to the rest of the graph as the modules they replace)
yields a \emph{modular decomposition (tree)} $T$.
Let $G_S$ denote the copy of $G$ in which all modules have been contracted to vertices.
Conveniently, each such $G_S$ admits a linear order $\prec_M$ over its maxcliques for some possible end $M \in \MC(G_S)$
and can thus be described up to isomorphism in $\C{k^*}{\bigO{\log n}}$ by \cref{lemm:stcc_to_c_k}.

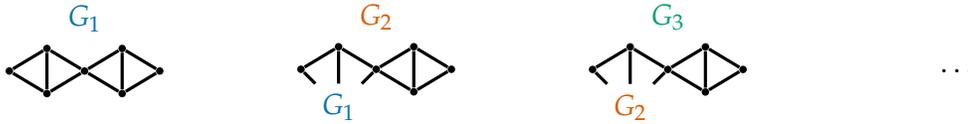
\begin{figure}
    \centering
    \begin{subfigure}[b]{.24\textwidth}
    \centering
        \begin{tikzpicture}
        \node[] (tag) at (1,.7) {$\textcolor{marine}{G_1}$};
        \node[circle,fill=black,inner sep=1pt] (1) at (0,0) {};
        \node[circle,fill=black,inner sep=1pt] (2) at (0.5,0.3) {};
        \node[circle,fill=black,inner sep=1pt] (3) at (0.5,-0.3) {};
        \node[circle,fill=black,inner sep=1pt] (4) at (1,0) {};
        \node[circle,fill=black,inner sep=1pt] (5) at (1.5,0.3) {};
        \node[circle,fill=black,inner sep=1pt] (6) at (1.5,-0.3) {};
        \node[circle,fill=black,inner sep=1pt] (7) at (2,0) {};
        \node[] (dummy) at (1,-0.65) {};

        \draw[-,very thick,draw=black] (1)  -- (2);
        \draw[-,very thick,draw=black] (1)  -- (3);
        \draw[-,very thick,draw=black] (2)  -- (3);
        \draw[-,very thick,draw=black] (2)  -- (4);
        \draw[-,very thick,draw=black] (3)  -- (4);
        \draw[-,very thick,draw=black] (4)  -- (5);
        \draw[-,very thick,draw=black] (4)  -- (6);
        \draw[-,very thick,draw=black] (5)  -- (6);
        \draw[-,very thick,draw=black] (5)  -- (7);
        \draw[-,very thick,draw=black] (6)  -- (7);
        \end{tikzpicture}
    \end{subfigure}
    \begin{subfigure}[b]{.24\textwidth}
    \centering
        \begin{tikzpicture}
        \node[] (tag) at (1,.7) {$\textcolor{red}{G_2}$};
        \node[circle,fill=black,inner sep=1pt] (1) at (0,0) {};
        \node[circle,fill=black,inner sep=1pt] (2) at (0.5,0.3) {};
        \node[] (3) at (0.5,-0.5) {$\textcolor{marine}{G_1}$};
        \node[circle,fill=black,inner sep=1pt] (4) at (1,0) {};
        \node[circle,fill=black,inner sep=1pt] (5) at (1.5,0.3) {};
        \node[circle,fill=black,inner sep=1pt] (6) at (1.5,-0.3) {};
        \node[circle,fill=black,inner sep=1pt] (7) at (2,0) {};

        \draw[-,very thick,draw=black] (1)  -- (2);
        \draw[-,very thick,draw=black] (1)  -- (3);
        \draw[-,very thick,draw=black] (2)  -- (3);
        \draw[-,very thick,draw=black] (2)  -- (4);
        \draw[-,very thick,draw=black] (3)  -- (4);
        \draw[-,very thick,draw=black] (4)  -- (5);
        \draw[-,very thick,draw=black] (4)  -- (6);
        \draw[-,very thick,draw=black] (5)  -- (6);
        \draw[-,very thick,draw=black] (5)  -- (7);
        \draw[-,very thick,draw=black] (6)  -- (7);
        \end{tikzpicture}
    \end{subfigure}
    \begin{subfigure}[b]{.24\textwidth}
    \centering
        \begin{tikzpicture}
        \node[] (tag) at (1,.7) {\textcolor{green}{$G_3$}};
        \node[circle,fill=black,inner sep=1pt] (1) at (0,0) {};
        \node[circle,fill=black,inner sep=1pt] (2) at (0.5,0.3) {};
        \node[] (3) at (0.5,-0.5) {\textcolor{red}{$G_2$}};
        \node[circle,fill=black,inner sep=1pt] (4) at (1,0) {};
        \node[circle,fill=black,inner sep=1pt] (5) at (1.5,0.3) {};
        \node[circle,fill=black,inner sep=1pt] (6) at (1.5,-0.3) {};
        \node[circle,fill=black,inner sep=1pt] (7) at (2,0) {};

        \draw[-,very thick,draw=black] (1)  -- (2);
        \draw[-,very thick,draw=black] (1)  -- (3);
        \draw[-,very thick,draw=black] (2)  -- (3);
        \draw[-,very thick,draw=black] (2)  -- (4);
        \draw[-,very thick,draw=black] (3)  -- (4);
        \draw[-,very thick,draw=black] (4)  -- (5);
        \draw[-,very thick,draw=black] (4)  -- (6);
        \draw[-,very thick,draw=black] (5)  -- (6);
        \draw[-,very thick,draw=black] (5)  -- (7);
        \draw[-,very thick,draw=black] (6)  -- (7);
        \end{tikzpicture}
    \end{subfigure}
    \begin{subfigure}[b]{.24\textwidth}
    \centering
        \begin{tikzpicture}
            \node[] (dot) at (1,1) {$\dots$};
            \node[] (dummy) at (1,0.35) {};
        \end{tikzpicture}
    \end{subfigure}
    \caption{First members of a family $(G_i)_{i\ge1}$ of interval graphs such that, for all $i \in \Nplus$, the modular decomposition of $G_i$ has a height of $i$. 
    This is because, for $G_{i, i \ge 2}$, a large fraction $\left(> \frac{|V(G)|}{2}\right)$ of its vertices is contained in its (unique) module.
    }
    \label{fig:interval_linear}
\end{figure}

Thus, for each $t\in V(T)$, the subgraph it corresponds to can be described in $\C{k^*}{\bigO{\log n}}$.
Hence, developing a formula which, in an inductive, bottom-up fashion, characterises $G$ up to isomorphism is not very difficult.
The main issue is the height of $T$.
This is due to the fact that an interval graph $G$ may have up to and at most one module $S_{\mathcal{C}}$ with $|S_{\mathcal{C}}| > |G|/2$.
Thus, we can construct families of interval graphs whose modular decomposition trees are linear in the size of the graph (see \cref{fig:interval_linear} for an example).

The idea is thus to build a treelike decomposition of the modular decomposition tree similarly as \cref{lem:5}.
Since this decomposition has a logarithmic height in the size of the input graph,
it then suffices to inductively describe our graph based on it, yielding formulae of logarithmic quantifier depth.
  
\section{Conclusion}
\label{sec:conclusion}
We have shown that for every \LRECeq-definable property,
there is a constant \(k\) such that for every size bound \(n\),
the property can be expressed on structures of size at most \(n\)
via a family of \(\logCk\)-formulae.
This implies that the $k$-dimensional Weisfeiler--Leman algorithm
distinguishes every pair of graphs separable by the property
in a logarithmic number of iterations.
Moreover, building on results by
Grohe et al.\ \cite{laubner2013recursion}
and Grußien~\cite{grussien2019capturing},
this yields that the algorithm identifies every interval graph
and every chordal claw-free graph in logarithmically many iterations.

It remains an interesting project to investigate the power of counting logics with logarithmic quantifier depth, or equivalently, a logarithmic number of iterations of the Weisfeiler--Leman algorithm, on other graph classes. A natural target class would be graphs defined by a finite set of excluded minors.

Also, since our results are non-uniform as we obtain formulae for each size bound $n$, a follow-up question could ask how to obtain similar uniform statements: is every $\LRECeq$-formula equivalent to a formula of fixed-point logic with counting that only uses logarithmically many iterations? 
\bibliographystyle{plainnat}
\bibliography{bibliography}

\end{document}